\documentclass[sigconf]{acmart}

\usepackage{algorithm}
\usepackage{algpseudocode}
\usepackage{amsmath}
\usepackage{amsthm}
\usepackage{minted}
\usepackage{graphicx}
\usepackage{subcaption}
\DeclareMathOperator*{\argmin}{argmin}

\usepackage{xcolor}
\usepackage{csvsimple}
\usepackage{balance}
\usepackage{multicol}
\usepackage{bbm}
\usepackage{enumitem}

\usepackage{hyperref}

\usepackage[svgnames]{xcolor}
\usepackage{tcolorbox}

\AtBeginDocument{%
  }

\setcopyright{acmlicensed}
\copyrightyear{2026}
\acmYear{2026}
\acmDOI{XXXXXXX.XXXXXXX}
\acmConference[Conference acronym 'XX]{Make sure to enter the correct
  conference title from your rights confirmation email}{June 03--05,
  2018}{Woodstock, NY}
\acmISBN{978-1-4503-XXXX-X/2018/06}

\newcommand{\one}[1]{{#1}}
\newcommand{\two}[1]{{#1}}
\newcommand{\three}[1]{{#1}}
\newcommand{\M}[1]{{#1}}

\setcopyright{none}
\begin{document}



\title{\textsc{ScaleDoc}: Scaling LLM-based Predicates over Large Document Collections}

\author{Hengrui Zhang}
\authornote{Both authors contributed equally to this research.}
\affiliation{%
  \institution{Tsinghua University}
    \city{Beijing}
  \country{China}
}
\email{zhanghen22@mails.tsinghua.edu.cn}

\author{Yulong Hui}
\authornotemark[1]
\affiliation{%
  \institution{Tsinghua University}
      \city{Beijing}
  \country{China}
}
\email{huiyl22@mails.tsinghua.edu.cn}

\author{Yihao Liu}
\affiliation{%
  \institution{Tsinghua University}
      \city{Beijing}
  \country{China}
}
\email{liuyihao24@mails.tsinghua.edu.cn}

\author{Huanchen Zhang}
\authornote{Huanchen Zhang is also affiliated with the Shanghai Qi Zhi Institute. Corresponding author.}
\affiliation{%
  \institution{Tsinghua University}
      \city{Beijing}
  \country{China}
}
\email{huanchen@tsinghua.edu.cn}


\begin{abstract}
Predicates are foundational components in data analysis systems. 
However, modern workloads increasingly involve unstructured documents, which demands semantic understanding, beyond traditional value-based predicates. 
Given enormous documents and ad hoc queries, while Large Language Models (LLMs) demonstrate powerful zero-shot capabilities, their high inference cost leads to unacceptable overhead.
Therefore, we introduce \textsc{ScaleDoc}, a novel system that addresses this by decoupling predicate execution into an offline representation phase and an optimized online phase. In the offline phase, \textsc{ScaleDoc} leverages a LLM to generate semantic representations for each document. Online, for each query, it adaptively trains a lightweight proxy model on these representations to filter the majority of documents, forwarding only the ambiguous cases to the LLM for final decision. Furthermore, \textsc{ScaleDoc} proposes two core innovations to achieve significant efficiency: (1) a contrastive-learning-based framework that trains the proxy model to generate reliable predicating decision scores; (2) an adaptive cascade mechanism that determines the effective filtering policy while meeting specific accuracy targets. Our evaluations across three datasets demonstrate that \textsc{ScaleDoc} achieves over a 2$\times$ end-to-end speedup and reduces expensive LLM invocations by up to 85\%, making large-scale semantic analysis practical and efficient.
\end{abstract}



 \maketitle

\section{Introduction}
\label{s-intro}

Predicates are essential for selecting relevant data in relational databases, search engines, and big-data systems.
Traditionally, these systems have excelled at value-based predicates (e.g., CITY = `New York').
However, modern analytical tasks increasingly require querying enormous corpora of unstructured documents based on their semantic meaning~\cite{huiuda,patel2024lotus}.
For example, medical researchers might look for all publications that ``developed novel psychotropic medications,'' or enterprises may need to analyze reports where customers are ``expressing dissatisfaction with service quality.''
Such queries require a deep contextual understanding that goes beyond simple keyword matching.

Handling these ad hoc semantic queries in large-scale analytics presents a significant challenge.
Conventional machine learning (ML) models are
not scalable for this purpose, because the extensive engineering
and data labeling effort is required for each new task.
While Large Language Models (LLMs) offer a zero-shot solution with their remarkable general capabilities, their high computational cost creates a major barrier.
The expense of running LLM inference on millions of documents for every query makes this approach impractical for widespread adoption~\cite{arora2023llm-dataview,hui2025okralong}.

Real-world semantic judgments are often ambiguous, making perfect accuracy costly and impractical.
Consequently, many systems prioritize scalable efficiency by aiming for a specific accuracy target~\cite{salazar2024inferdb,kang2017noscope,yang2022optimizing,lu2018accelerating}.
To achieve this trade-off, some systems, such as NoScope~\cite{kang2017noscope} and PPs~\cite{lu2018accelerating}, use lightweight proxy models to filter easier cases.
\one{
However, these proxies are designed for traditional small models and specific tasks. They lack zero-shot flexibility and fail to bridge the vast capability gap with massive LLMs.}
More recent LLM-centric solutions, such as FrugalGPT~\cite{chen2023frugalgptuselargelanguage} and LOTUS~\cite{10.14778/3749646.3749685}, use smaller LLMs (e.g., GPT-3.5 or LLaMA-8B) as filtering proxies.
While more flexible, these smaller LLMs remain too computationally expensive for truly large-scale applications involving millions of documents.

A critical inefficiency in LLM-based approaches is the repetitive re-processing of entire documents for every ad hoc query, incurring substantial and redundant computational costs.
A key insight is to shift these expensive, document-centric LLM computations to a one-time offline phase to reduce the burden during online processing.
We, therefore, propose \textsc{ScaleDoc}, a system designed for efficient LLM-based predicate execution over enormous document corpora.
\textsc{ScaleDoc} decouples the execution into two phases.
The one-time \emph{offline representation} phase creates a rich semantic representation of each document using an LLM.
Then, when an ad-hoc query arrives, the \emph{online processing} phase trains a lightweight, query-specific proxy model that uses these representations to rapidly filter the documents.
The model identifies and forwards only the most ambiguous documents to the powerful but expensive LLM, achieving significant efficiency gains while maintaining accuracy.

\one{While this architecture is promising, basic engineering integration is insufficient. Basic approaches often fail to address the fundamental capacity mismatch, inherent in scaling LLMs for complex scenarios.}
Therefore, the effectiveness of \textsc{ScaleDoc} depends on overcoming two key challenges. First, the lightweight proxy must be trained to provide reliable \textbf{decision scores}. Lacking the oracle LLM's deep semantic understanding, standard training methods often yield ambiguous scores that fail to distinguish positive from negative cases. Under such proxy ambiguity, most documents would still be forwarded to the expensive oracle. \one{To address this, we propose a new design principle: \textbf{query-aware distribution shaping}}. Guided by this principle, we developed a framework based on {contrastive learning} that trains the proxy to capture fine-grained semantics and produce well-behaved predicating scores.

The second challenge is that the system must determine an effective \textbf{filtering criterion} for each ad hoc query.
Because the relationship between indeed accuracy and decision scores is unknown for a new query, it is difficult to establish a filtering threshold that meets a user's accuracy target while minimizing cost.
We tackled this with an \textbf{adaptive cascade mechanism} that performs online calibration to shape query-specific score distributions, and then uses an optimized algorithm to determine the ideal filtering thresholds.

We evaluated \textsc{ScaleDoc} across three diverse datasets, where it significantly outperforms existing baselines.
On average, \textsc{ScaleDoc} achieves over 2$\times$ end-to-end performance speedup and reduces costly LLM invocations by up to 85\%.

Our paper makes the following contributions:
\begin{itemize}
    \item We propose \textsc{ScaleDoc}, a novel system that decouples LLM execution into a one-time offline representation phase and an optimized online query phase.
    \item We introduce a contrastive learning strategy to train a lightweight, query-aware proxy model that provides reliable predicating decision scores.
    \item We design an adaptive online calibration mechanism with an optimized filtering algorithm to guarantee user-specified accuracy while minimizing calls to the expensive oracle LLM.
\end{itemize}

\begin{figure*}[t!]
\centering
  \includegraphics[trim=0cm 4.5cm 1cm 4.5cm, clip, width=0.85\linewidth]{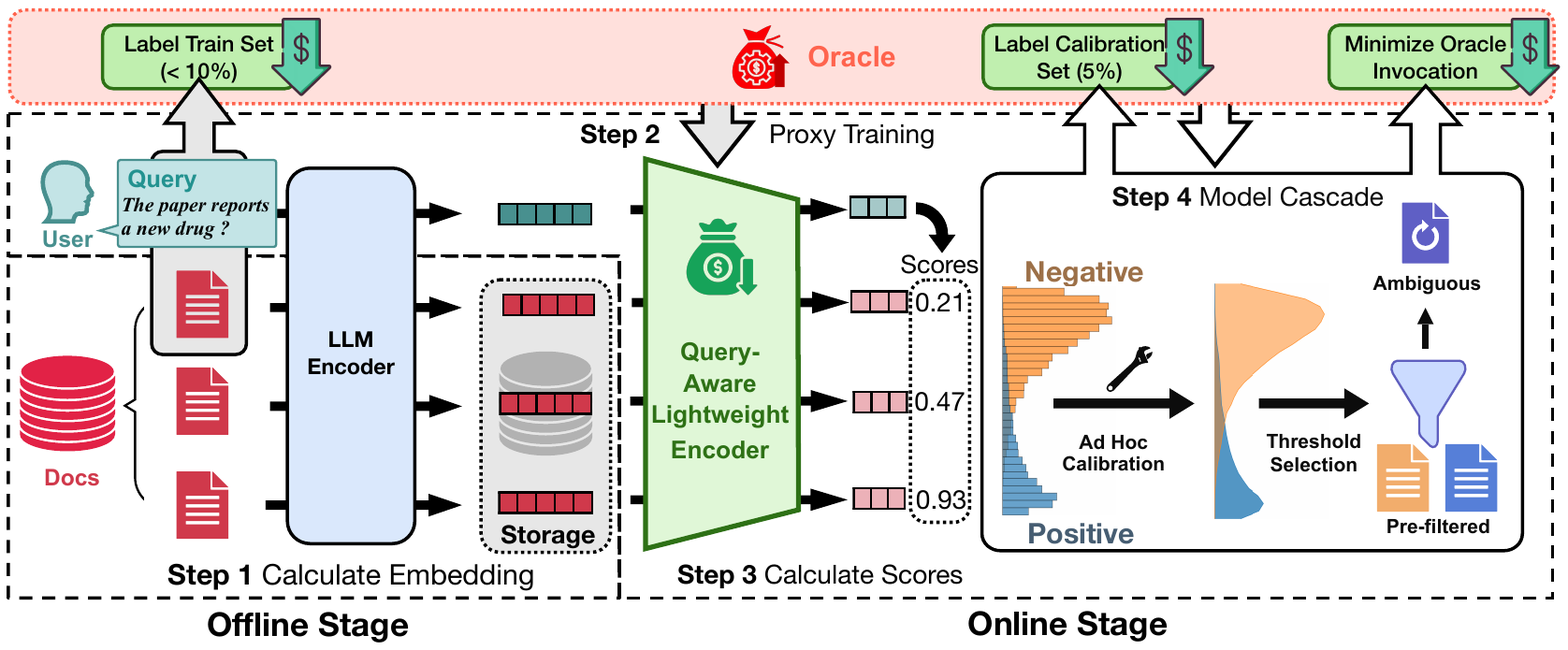}
  \caption{A detailed workflow of \textsc{ScaleDoc} -- \textmd{\textsc{ScaleDoc}  adapts pre-calculated semantic embeddings for query-specific online processing. The online process comprises a query-aware lightweight encoder and a subsequent cascade workflow.}}
  \label{fig:teaser}
\end{figure*}

\section{System Overview}

\textsc{ScaleDoc} provides an efficient solution for executing semantic predicates over large-scale documents. This section defines the target workloads, introduces the system's architecture, and outlines the core challenges to overcome.

\subsection{Workload Specification}
A semantic predicating workload consists of a large document collection $D$, and a set of queries $Q$. Each query, $q \in Q$, is defined by two components: a natural language predicate and a user-specified accuracy target.
Formally, we use an SQL-like syntax to express these queries. For instance, a query to find all medical papers in PubMed that introduce a new drug would be:

\begin{minted}{sql}
SELECT * FROM PubMed
WHERE "The paper introduces a new drug"
WITH accuracy_target = 0.90
\end{minted}
For each such query, the final task is to evaluate the predicate against every document $d \in D$ and assign a binary label (positive or negative), while meeting the specified accuracy target.

\subsection{System Architecture}

\textsc{ScaleDoc} adopts an offline-online architecture, enabling efficient execution by decoupling query processing stages. An overview of the \textsc{ScaleDoc} pipeline is illustrated in \autoref{fig:teaser}.

The offline stage is a one-time, compute-intensive process. For each document, we use a small-scale LLM (e.g. with 7B parameters) to generate a semantic embedding, which is then stored for online use. This approach offers two main benefits. First, it leverages the expressive power of the LLM to produce semantically rich document representations. This provides a high-quality foundation, allowing lightweight processing in subsequent online stages. Second, it front-loads the necessary yet expensive computations of LLM. By pre-computing and storing these embeddings,
\textsc{ScaleDoc} can efficiently reuse them across countless ad hoc queries, eliminating the need to repeatedly run the LLM.

The online stage employs a proxy-cascade architecture to handle ad hoc queries efficiently.
Upon receiving a new query, \textsc{ScaleDoc} will train a query-specific lightweight proxy model. This model then rapidly evaluates the documents, assigning each a decision score, indicating the likelihood of the positive predicate. To train this model, we need to first sample a small fraction (e.g. $5\%$) of the documents, and obtain ground-truth predicating labels by calling a powerful oracle LLM (e.g. GPT-4o ~\cite{openai2024gpt4ocard}). After the proxy model, a subsequent cascade filter uses the proxy decision scores to decide the high-confidence and low-confidence documents. The proxy's judgment is adopted for the high-confidence set, whereas the low-confidence set is forwarded to the oracle LLM for a final judgment. This hierarchical approach strategically minimizes oracle invocations, ensuring scalable and cost-effective query execution.

\subsection{Core Challenges}
While this offline-online and proxy-oracle architecture is promising, its effectiveness hinges on overcoming two critical challenges:

\textbf{Reliable Proxy Model.}\hspace{0.2cm} The efficiency of the cascade filter depends on the quality of the decision scores produced by the lightweight proxy model. First, a naive model might fail to capture the fine-grained semantics between the query and documents, leading to unreliable proxy predictions. Second,
a poorly trained model may generate ambiguous scores that do not clearly distinguish positive and negative documents. Such an ambiguity would force a large fraction of instances to be categorized as ``uncertain'', leading to excessive invocations of the expensive oracle LLM and diminishing the system's performance. Therefore, the first challenge is to develop a well-behaved proxy model, capable of generating accurate and decisive scores, enabling maximum data reduction.

\textbf{Ad Hoc Cascade.}\hspace{0.2cm}
Given the proxy model's scores, the cascade filter must determine the decision thresholds to separate documents into high-confidence (to be filtered) and low-confidence (to be verified by the oracle) sets. The goal is to meet the user-specified accuracy target while minimizing the number of oracle calls. However, in an ad hoc setting, neither the query-specific data distribution nor the ground-truth labels are known beforehand. This lack of prior knowledge makes it non-trivial to select effective thresholds online. Therefore, the second challenge is to design an efficient online calibration mechanism that can dynamically determine the effective thresholds, ensuring the final accuracy target is met with minimal oracle overhead.

To address these two challenges, \textsc{ScaleDoc} introduces a novel contrastive-learning-based approach to train a robust proxy model (Section ~\ref{s4}) and proposes an efficient calibration workflow for the model cascade (Section ~\ref{s5}). Together, these contributions ensure both high data reduction and robust predicating accuracy.

\section{Query-Aware  Model Training}
\label{s4}

The online efficiency of \textsc{ScaleDoc} hinges on a lightweight, query-aware proxy model. For each ad hoc query, this model is trained to emulate the semantic judgments of the LLM oracle. It takes precomputed document representations as input and outputs a decision score for each document, indicating the likelihood of satisfying the semantic predicate.  This scoring enables a cascaded filtering process: high-confidence positive and negative documents are filtered, while only low-confidence, ambiguous documents are sent to the LLM oracle. Therefore, these decision scores, used to reduce oracle invocations, are paramount to overall system performance.

However, training an effective lightweight proxy model is non-trivial. Basic approaches, such as the naive binary classifiers in prior work~\cite{lu2018accelerating, yang2022optimizing}, struggle to capture query-specific semantics. Our preliminary experiments with a more advanced query-fused regression MLP also yielded unreliable scores.

To address this, we argue that the training process must be explicitly designed to produce scores with a well-behaved distribution. We achieve this through a contrastive learning framework. In Section~\ref{s4.1}, we detail the desirable properties of this score distribution. Section~\ref{s4.2} then describes how \textsc{ScaleDoc} leverages our contrastive learning mechanism to generate the scores.




\subsection{Desirable Distribution on Decision Scores}
\label{s4.1}

The primary goal of the lightweight model is data reduction, filtering more documents based on their decision scores. For clarity, we assume higher scores indicate a higher confidence in being positive and lower scores indicate being negative. 
Therefore, the low-score and high-score thresholds determine the filtering.

In practice, a high data reduction rate is not always guaranteed. \autoref{fig:bad} illustrates how a poorly structured score distribution (\textit{logprobs} of Llama-3.2-3B ~\cite{grattafiori2024llama3herdmodels}'s first response token) can severely hinder data reduction. Specifically, the real positive class exhibits two peaks, appearing in both high and low scores. 
The attempt to set a filter threshold in the low-score region (e.g., the left red line) would \textbf{misclassify} many positives, violating the accuracy target and forcing a conservative and ineffective threshold.
In contrast, \autoref{fig:good} illustrates a desirable distribution with a clear separation. Only minimal positive items fall below the red line threshold, allowing effective filtering of the true negatives.

To consistently achieve effective filtering, we propose three essential properties for the decision scores: 
\vspace{1pt}

\begin{figure}[t]
  \centering
  \begin{subfigure}[b]{0.45\linewidth}
    \includegraphics[width=\linewidth]{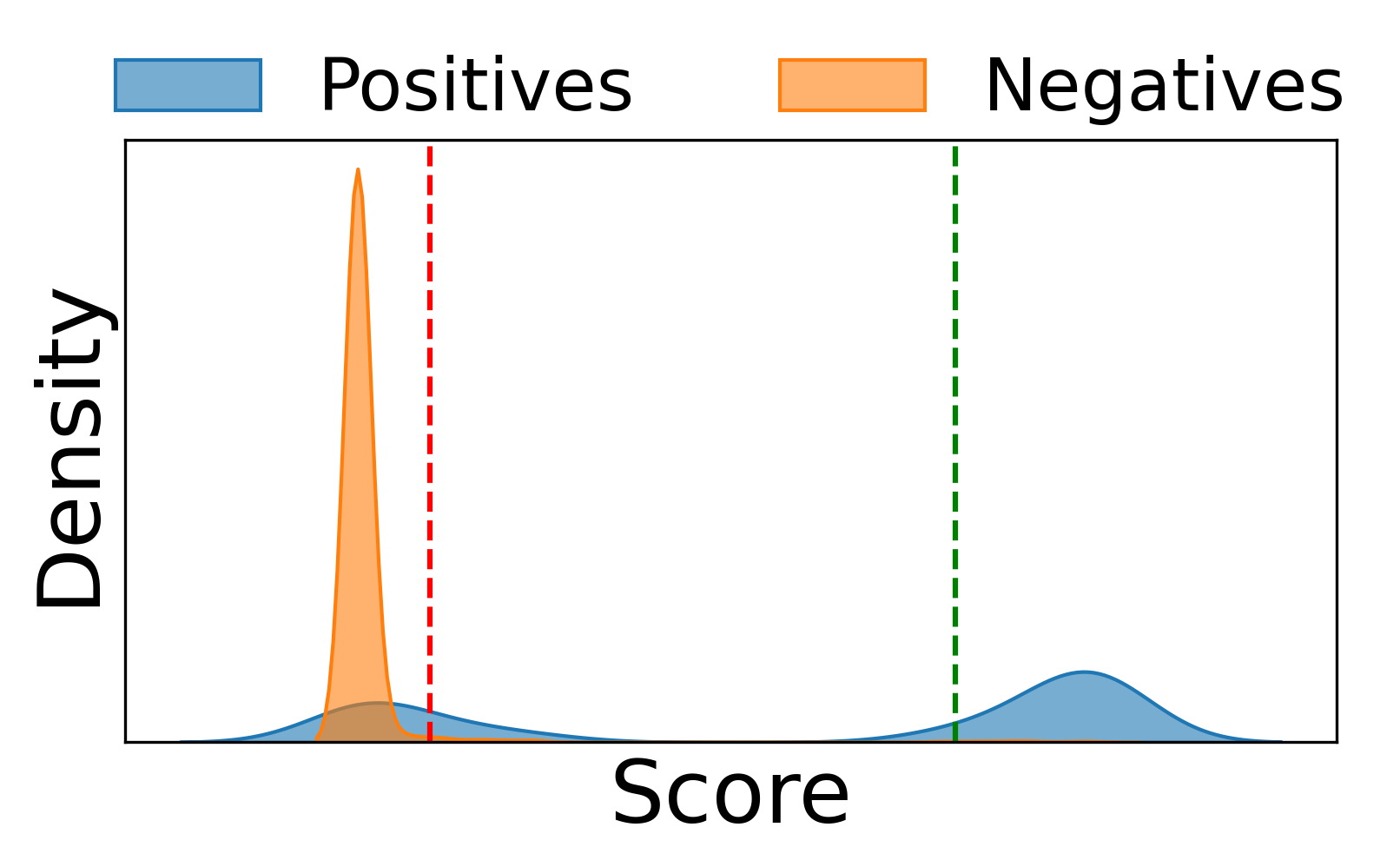}
    \caption{Llama-3.2-3B logprobs}
    \label{fig:bad}
  \end{subfigure}
  \hfill
  \begin{subfigure}[b]{0.45\linewidth}
    \includegraphics[width=\linewidth]{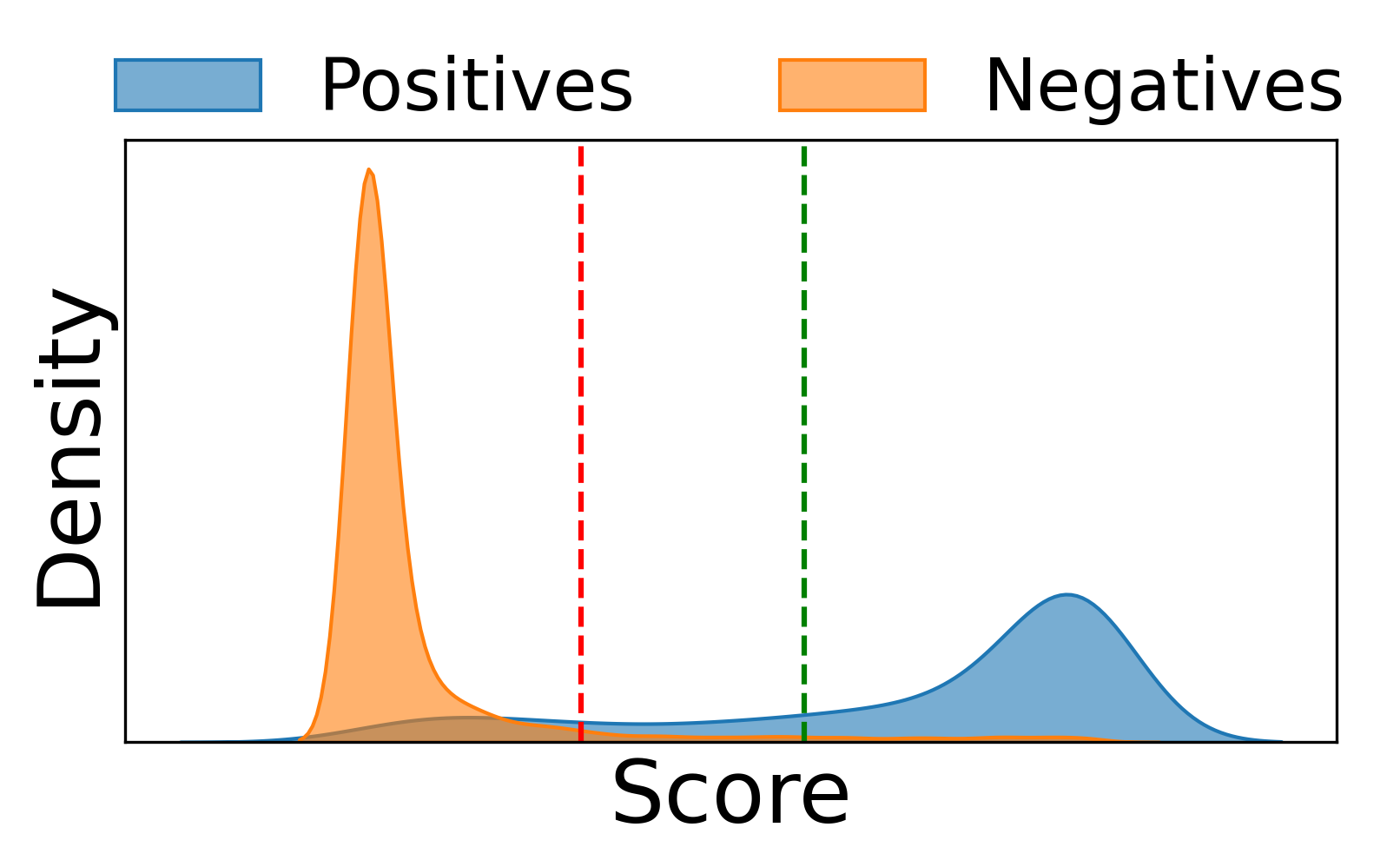}
    \caption{ScaleDoc}
    \label{fig:good}
  \end{subfigure}
  
  \caption{Example score distributions of different proxies, with low and high data reduction rate.}
  \label{fig:distr}
\end{figure} 

\textbf{1) Smoothness.} The distribution of decision scores should be continuous and smooth, without abruptions or discontinuities. This ensures the threshold-selection algorithms can operate stably and consistently.
We have observed that this fundamental property fails with some naive proxies such as Kernel Density Estimation (KDE).

\textbf{2) Semantic monotonicity.} A document that is more semantically satisfying the query should be assigned a higher decision score, and vice versa. This prevents reward-hacking of deterministic score ranges, ensuring a meaningful and reliable ranking. 

\textbf{3) Bipolarity.} The score distribution must be strongly polarized, with positive and negative documents clustering distinctly at the high and low ends of the score spectrum.
This clear separation is the primary driver of high data reduction. 

\vspace{1pt}

Achieving these desirable properties is challenging. While LLM-based embeddings offer rich semantics, they are static and not optimized for the specific semantic distinctions of an ad-hoc query. Directly training a lightweight model (e.g., MLP) on these embeddings typically yields inconsistent score distributions. To overcome this and derive discriminative query-specific patterns, we employ a contrastive learning-based approach, which excels at capturing fine-grained, task-specific distinctions. 

\subsection{A Contrastive Learning-based Approach}
\label{s4.2}

\textsc{ScaleDoc} employs a contrastive-learning-based framework, designed to refine pre-computed embeddings into tailored and semantic-aligned decision scores.
The core is a \textbf{lightweight encoder} $E(\cdot)$, which utilizes the MLP structure, mapping both the document and query into a shared \textbf{latent space}.  Following this, the final decision score is the cosine similarity between the encoded representations of the query and documents. This score is inherently smooth, and through our tailored training, optimized to be semantically monotonic and bipolar.

Formally, given a query $q$  and a document $d$, we first obtain their high-dimensional semantic embeddings from an LLM encoder, denoted as $\mathbf{e}_q \in \mathbb{R}^D$ and $\mathbf{e}_d \in \mathbb{R}^D$. \textsc{ScaleDoc} then employs a lightweight encoder, $E(\cdot): \mathbb{R}^D \to \mathbb{R}^l$, mapping these embeddings into a latent space $\mathbb{R}^l$:
$$\mathbf{z}_q = E(\mathbf{e}_q), \quad \mathbf{z}_d = E(\mathbf{e}_d)$$

The decision score, $s(q, d)$, for the query-document pair is defined as the cosine similarity between their latent representations:
$$s(q, d) = \text{sim}(\mathbf{z}_q, \mathbf{z}_d) = \frac{\mathbf{z}_q \cdot \mathbf{z}_d}{\|\mathbf{z}_q\| \|\mathbf{z}_d\|}$$

For this task $T$, consisting of a query $q$ and a collection of documents $\mathcal{D} = \{d_1, d_2, \dots, d_N\}$, the complete set of decision scores can be expressed as:
$$S(T) = S(q, \mathcal{D}) = \{s(q, d_i) \mid d_i \in \mathcal{D}\}$$


\begin{figure*}
      \centering
  \includegraphics[trim=1cm 8cm 1cm 5cm, clip, width=0.8\linewidth]{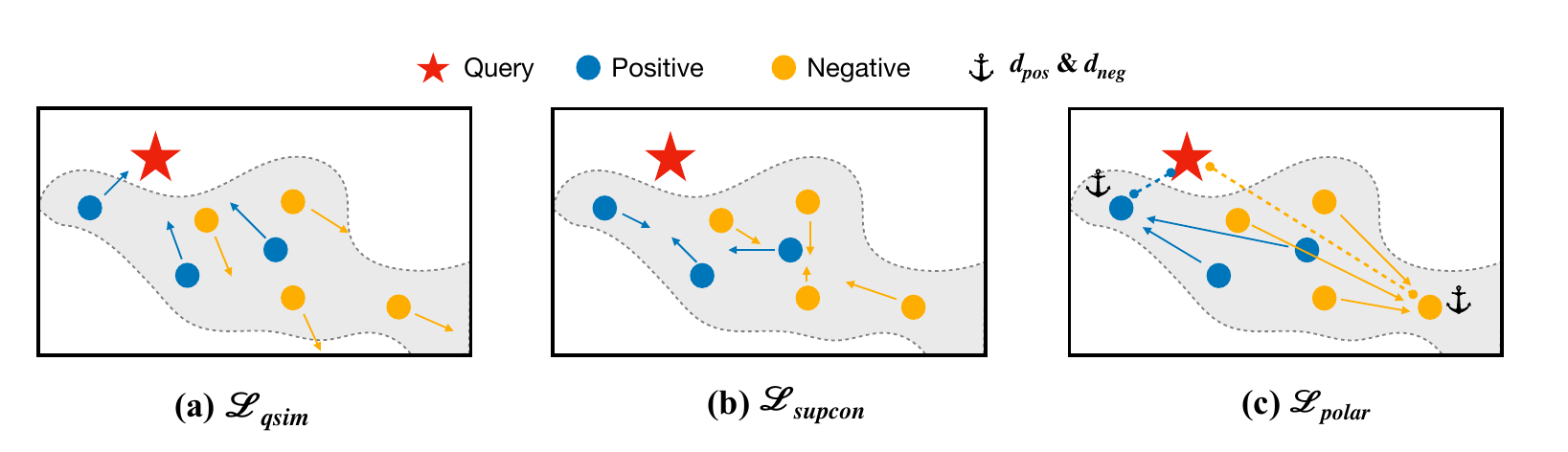}
  \caption{Illustration of the objectives adopted in training \textsc{ScaleDoc}'s Query-Aware Encoder.}
  \label{fig:loss}
\end{figure*}

To effectively shape the latent space and meet the required properties, \textsc{ScaleDoc} trains the encoder in two distinct phases. This two-phase strategy is crucial because jointly optimizing all properties simultaneously can lead to conflicting training signals and results in observed performance decline. Phase 1 first establishes the correct semantic relationship as the foundation. Then Phase 2 refines the spectrum of the embedding space to create a clear separation between positive and negative cases. There are three objectives during the overall training process, depicted in \autoref{fig:loss}.

\vspace{4 pt}
\textbf{Training Phase 1: Semantic Monotonicity}
\hspace{0.2cm} This phase aims to build the foundational semantic relationship between the documents and the query. 
We use a contrastive loss $\mathcal{L}_{qsim}$, inspired by dense passage retrieval ~\cite{karpukhin2020dense}. In our training, the query embedding $\mathbf{z}_q$ acts as an anchor. The objective is to pull positive document embeddings ($d^+$) closer to the anchor while pushing negative ones ($d^-$) away in the latent space, as shown in \autoref{fig:loss}(a).

Formally, given a query $q$, each input mini-batch $\mathcal{D'}$ comprises $m$ positive documents $\{d^+_i\}_{i=1}^m$ and $n-m$ negative documents $\{d^-_j\}_{j=1}^{n-m}$, hence $\mathcal{D'} =\{d_i^+\}\bigcup\{d_j^-\}$. The training loss is defined as: 
\begin{equation}
    \mathcal{L}_{qsim}(q, \mathcal{D'}) =
    -log\frac{\sum\limits_{i=1}^{m}{e^{\text{sim}(q, d^+_i)/\tau}}}{\sum\limits_{d \in \mathcal{D'}}{e^{\text{sim}(q, d)/\tau}}}
\end{equation}
Here, $\tau$ is a temperature hyperparameter. 
The objective utilizes negative log-likelihood to maximize similarity between queries and positive documents, while penalizing negatives. This relocates positives closer to the query within the latent space.

A crucial distinction from previous work ~\cite{karpukhin2020dense} is that we train our encoder \textbf{dynamically} for each new query task. This allows the model to learn query-specific semantics rather than a single static matching function. It is essential in our setting: our online proxy model is intentionally quite lightweight for efficiency, lacking the general capacity of large models. By specializing the proxy for each query, we enable it to perform effectively despite its limited size.

However, this training phase alone is insufficient to guarantee the robust bipolarity. 
Without an explicit separation margin, we observe that local neighbors (mixed positive and negative cases) often relocate together.
This results in distribution overlap and class collapse, which hinders effective data filtering.

\vspace{4 pt}
\textbf{Training Phase 2: Enforcing Bipolarity} \hspace{0.2cm} 
To address the limitations of Phase 1, the second phase explicitly shapes the latent space to form a bipolar distribution. Using the same encoder and training data, we
introduce two loss functions, $\mathcal{L}_{supcon}$ and $\mathcal{L}_{polar}$.  

First,  $\mathcal{L}_{supcon}$ adopts supervised-contrastive-learning to encourage intra-class clustering ~\cite{khosla2020supervised}, pulling documents of the same label together, as shown in \autoref{fig:loss}(b). 
The loss is formulated as:
\begin{equation}
    \mathcal{L}_{supcon}(\mathcal{D'}) = 
    -\sum_{i=1}^{n}\frac{1}{|U(i)|}\log\frac{\sum_{d_p \in U(i)}e^{\text{sim}(d_i, d_p)/\tau}}{\sum_{d_k \in A(i)}e^{\text{sim}(d_i, d_k)/\tau}}
    \label{eq:l_supcon}
\end{equation}
Within a mini-batch, for each anchor document $d_i$ , $U(i)$ denotes documents
sharing the same label with $d_i$, and $A(i)$ comprises all other documents . 
With this objective, we can create a discriminative embedding space that minimizes intra-class variance, crucial for the proxy to learn a clear decision boundary.


Second, $\mathcal{L}_{polar}$ introduces a novel mechanism to explicitly enforce a bipolar manifold and enlarge the margin between classes. 
The intuition of $\mathcal{L}_{polar}$ is selecting a \textit{bellwether} sample for positives and negatives in each mini-batch to guide the direction of clustering. 
A bellwether is distinguished by the closest (positive) or furthest (negative) distances relative to the query inside each mini-batch. For a given mini-batch and query, we define the bellwethers as:
$$d_{pos} = \underset{d_i \in \{d^+\}}{\text{argmin}} \hspace{0.1cm} \text{sim}(q, d_i), \quad d_{neg} = \underset{d_j \in \{d^-\}}{\text{argmax}} \hspace{0.1cm} \text{sim}(q, d_j)$$
$\mathcal{L}_{polar}$ then uses these bellwethers as anchors, pulling positive documents towards $d_{pos}$ and and negative documents towards $d_{neg}$.


\begin{multline}
    \mathcal{L}_{polar} (q, \mathcal{D'}) = \\
    - log\frac{\sum\limits_{i=1}^{m}{e^{sim(d_{pos}, d^+_i)/\tau}}}{\sum\limits_{d\in \mathcal{D'}}{e^{sim(d_{pos}, d)/\tau}}} -
    log\frac{\sum\limits_{j=1}^{n-m}{e^{sim(d_{neg}, d^-_j)/\tau}}}{\sum\limits_{d\in\mathcal{D'}}{e^{sim(d_{neg}, d)/\tau}}} 
\end{multline}

Bellwethers can be different across mini-batches, but $\mathcal{L}_{polar}$ ensures a consistent relocation. The progressive process explicitly enlarges the separation margin between the positive and negative poles. This constructs a highly bipolar manifold, enhancing the efficacy of threshold-based filtering, as shown in \autoref{fig:loss}(c).


Collectively, these two training phases and three loss functions effectively shape the latent space to produce effective decision scores, satisfying our proposed properties. For the training data, we heuristically sample a small subset of documents (e.g., 5\%) and use the oracle LLM to generate ground-truth labels (more implementation details in Section \ref{s6}).

\section{Model Cascade}
\label{s5}

After the proxy procedure, \textsc{ScaleDoc} uses a powerful but expensive LLM oracle to resolve uncertain documents through the cascade component. Model cascade aims to meet the user-specified accuracy while minimizing oracle invocations by filtering most documents and only forwarding the ambiguous cases to the oracle.

In ad hoc settings, the lack of prior knowledge about query-specific labels and data distribution poses a significant challenge, often leading to suboptimal data reduction in existing methods~\cite{10.14778/3407790.3407804}. \textsc{ScaleDoc} addresses this with a novel calibration workflow and an optimized filtering algorithm. Leveraging the well-behaved decision scores from our lightweight model (Section \ref{s4}), \textsc{ScaleDoc} achieves superior data reduction while robustly achieving accuracy targets.

This section proceeds as follows. Section \ref{s5.1} formulates the cascade task as a constrained optimization problem. Section \ref{s5.2} introduces our ad hoc calibration method, which enables accurate performance estimation from a small sampled subset. Section \ref{s5.3} details the algorithm for selecting optimal filtering thresholds.

\subsection{Problem Formulation}
\label{s5.1}
For a given query task $T$, the lightweight proxy model produces decision scores for all documents, denoted by $S(T)$. Each score is a cosine similarity within the interval $[0, 1]$, where the higher one indicates stronger semantic agreement with the query.  To filter out  high-confidence instances, the cascade component should select two thresholds, a lower bound $l$ and an upper bound $r$. The filtering logic is as follows:

\begin{itemize}
    \item Documents with scores in $(r, 1]$ are classified as \textbf{positive}.
    \item Documents with scores in $[0, l)$ are classified as \textbf{negative}.
    \item Documents with scores in $[l, r]$ are deemed \textbf{ambiguous} and are sent to the LLM oracle for final decision.
\end{itemize}


The primary goal is to minimize the fraction of documents sent to the LLM oracle, while satisfying the user-specified accuracy target $\alpha$. We define this fraction as the unfiltered rate $u$, controlled by the lower bound and upper bound:
\begin{equation}
u(l, r) = \frac{|\{s_i \in S(T) \mid l \le s_i \le r\}|}{|S(T)|}
\label{eq:unfilted_rate}
\end{equation}

The final accuracy, \textsc{Acc}$(l, r)$, is determined by the correctness of the lightweight model in the filtered regions, combined with the perfect accuracy of the oracle on the unfiltered region $[l, r]$.

This leads to the following optimization problem, where $\mathcal{K}$ represents the search space for $(l, r)$, with $0 \le l < r \le 1$.
\begin{equation}
\min_{(l, r) \in \mathcal{K}} u(l, r) \quad \text{s.t.} \; \textsc{Acc}(l, r) \ge \alpha
\label{eq:optimization}
\end{equation}

Solving this problem in an ad hoc setting is challenging, because the function \textsc{Acc}$(l, r)$ is unknown without full ground truth statistics. Therefore, a calibration process is required to estimate the accuracy, followed by the thresholds selection algorithm.

\subsection{Ad-hoc Calibration}
\label{s5.2}
To solve the optimization problem in ~\eqref{eq:optimization}, we must obtain the accuracy $\textsc{Acc}(l, r)$ for any given threshold pair $(l,r)$. However, in the ad hoc setting, the ground-truth labels for the full collection are not available. Therefore, \textsc{ScaleDoc} relies on a small oracle-labeled sample to estimate the accuracy and calibrate the cascade.

However, this poses a critical challenge: small samples may fail to capture the true global relationship between labels and proxy scores.  The sampled score distributions can deviate substantially from the true global distributions, which may lead to poorly calibrated threshold $(l,r)$ and fail to meet the  accuracy target.
This issue stems from two key limitations: (1) simple random sampling might under-represent documents in low-density regions and cause information loss ~\cite{10.14778/3407790.3407804}; (2) sample randomness may introduce stochastic noise.
While a sufficiently large sample could mitigate the estimation bias, the high cost of oracle labeling makes this infeasible.

To address this, \textsc{ScaleDoc} introduces a robust calibration workflow (Algorithm \ref{alg:calibr}), which reconstructs the global score distributions from a small sample. It comprises two main stages: \textbf{stratified sampling} and \textbf{distribution reconstruction}.
 Our experiments demonstrate that with a modest sample size of 5\%, \textsc{ScaleDoc} can achieve great cascade performance.

\begin{algorithm}[t]
\caption{Ad hoc Calibration Workflow}\label{alg:calibr}
\begin{algorithmic}[1]
\State\textbf{Given:} Decision scores $S(T)$, Oracle LLM
\State $bins \gets$ Discretize($S(T)$) \Comment{Discretize score range into bins for stratified sampling and analysis}
\State $S'(T) \gets$ StratifiedSample($S(T)$) \Comment{Sample representatively from each bin}
\State $S'_{P} \gets \{s_i \mid s_i \in S'(T) \land \text{Oracle}(q,d_i) = \text{positive}\}$
\State $S'_{N} \gets \{s_i \mid s_i \in S'(T) \land \text{Oracle}(q,d_i) = \text{negative}\}$
\State $PDF_P \gets \text{Smooth}(\text{DE}(\text{Jitter}(S'_{P})))$ \Comment{Reconstruct score distribution for positive class }
\State $PDF_N \gets \text{Smooth}(\text{DE}(\text{Jitter}(S'_{N})))$ \Comment{Reconstruct score distribution for negative class }
\State\textbf{return} $PDF_P, PDF_N, bins$
\end{algorithmic}
\end{algorithm}


\vspace{4pt}
\noindent\textbf{Stratified Calibration Sampling.} \hspace{0.2cm}
As stated above, uniform random sampling may fail to capture low-density score regions and introduce noise, especially with smaller sample sizes.
To address these issues, we propose to employ stratified sampling. We discretize the entire score range into a series of bins. Then, we sample from each bin in proportion to its population in the global set $S(T)$. This ensures that the sampled subset $S'(T)$ preserves the relative density of the global distribution, preventing low-density regions from being entirely omitted.

\vspace{4pt}
\noindent\textbf{Distribution Reconstruction.} \hspace{0.2cm}
After obtaining the labeled sample $S'(T)$, we reconstruct continuous and robust score distributions for positive and negative instances. The core objective is to faithfully mirror the global distributions $S(T)$, optimized through the following mathematical process:

\textbf{1) Jittering for Information Recovery.} \;
To counteract information loss, especially in low-density regions, we first apply random jittering. Stratified sampling with a modest rate can cause bins with few samples to become empty. Ignoring these gaps encourages overconfidence in these score ranges, creating a risk factor for subsequent threshold selection. Jittering addresses this by introducing low-density, random data into these empty bins, ensuring that this information is not lost before subsequent modeling.

\textbf{2) Density Estimation (DE) via Linear Interpolation.} \;
Next, we construct the continuous Probability Density Functions ($PDF_P$ and $PDF_N$) from the discrete jittered samples. We employ \textit{linear interpolation} for this density estimation task, providing virtually distortion-free modeling.
This process yields a continuous probabilistic interface that facilitates subsequent accuracy calculations via density integration rather than discrete sample counting. This allows querying over arbitrary scores for an estimated density value. Compared to methods like Kernel Density Estimation (KDE), linear interpolation provides a more faithful representation of the empirical distribution, particularly in low-density regions.

\textbf{3) Smoothing.} \;
Finally, to mitigate anomalous noise and information loss from sampling randomness, we apply the Moving Average (MA) approach. MA is a simple soft computing method that smooths the distribution by applying mean pooling within a fixed-size sliding window, which allows for a better capture of the general distributional features. Applying an MA filter reduces spikes and breakpoints, providing an additional layer of smoothing to reveal the underlying global trend more effectively.

With the above optimizations, the reconstructed distributions serve as a continuous and robust foundation for the subsequent threshold selection process. The final outputs of Algorithm \autoref{alg:calibr} are the distributions of positives and negatives in the form of PDFs, and the discretization of the score range \textit{bins}.


\subsection{Threshold Selection}
\label{s5.3}


\begin{algorithm}[t]
\caption{Thresholds Selection}\label{alg:threshold-selection}
\begin{algorithmic}[1]
\State\textbf{Given} Reconstructed $\mathit{CDF}_P, \mathit{CDF}_N$; discretization $\textit{bins}$
\State $\textit{steps} \gets \textit{bins.edges}$ \Comment{Initialize thresholds search space}
\State $l_s, r_s \gets \textit{steps}.\text{first}, \textit{steps}.\text{last}$
\vspace{0.2cm}



\State \textbf{function} \textsc{SelectThresholds} ($\alpha$)
\State\hspace{\algorithmicindent}$l_0, r_0 \gets l_s, r_s$ \Comment{Initialize thresholds}

\vspace{0.2cm}
\Comment{1. Find the tightest $l_0$ with $r=r_s$}
\State\hspace{\algorithmicindent}\textbf{for} $l$ \textbf{from} $\textit{steps.first}$ \textbf{to} $\textit{steps.last}$ \textbf{do}
\State\hspace{\algorithmicindent}\hspace{\algorithmicindent}\textbf{if} \textsc{Acc}$(l, r_s) \ge \alpha$ \textbf{then} $l_0 \gets l$ \textbf{else break}
\State\hspace{\algorithmicindent}\textbf{end for}

\vspace{0.2cm}
\Comment{2. Find the tightest $r_0$ with $l=l_s$}
\State\hspace{\algorithmicindent}\textbf{for} $r$ \textbf{from} $\textit{steps.last}$ \textbf{to} $\textit{steps.first}$ \textbf{do}
\State\hspace{\algorithmicindent}\hspace{\algorithmicindent}\textbf{if} \textsc{Acc}$(l_s, r) \ge \alpha$ \textbf{then} $r_0 \gets r$ \textbf{else break}
\State\hspace{\algorithmicindent}\textbf{end for}

\vspace{0.2cm}
\Comment{3. Construct frontier path $P$ from $(l_0, r_s)$ to $(l_s, r_0)$}
\State\hspace{\algorithmicindent}$l, r\gets l_0, r_s$
\State\hspace{\algorithmicindent}$P\gets \{(l_0, r_s)\}$
\State\hspace{\algorithmicindent}\textbf{while} $(l, r) \ne (l_s, r_0)$
\State\hspace{\algorithmicindent}\hspace{\algorithmicindent}$l_{next}, r_{next} \gets l + bins.size, r - bins.size$
\State\hspace{\algorithmicindent}\hspace{\algorithmicindent}\textbf{if} \textsc{Acc}$(l_{next}, r) \ge \alpha$ \textbf{then}
\State\hspace{\algorithmicindent}\hspace{\algorithmicindent}\hspace{\algorithmicindent}$l \gets l_{next}$
\State\hspace{\algorithmicindent}\hspace{\algorithmicindent}\textbf{else}
\State\hspace{\algorithmicindent}\hspace{\algorithmicindent}\hspace{\algorithmicindent}$r \gets r_{next}$
\State\hspace{\algorithmicindent}\hspace{\algorithmicindent}$P \gets P \bigcup \{(l, r)\}$
\vspace{0.1cm}
\Statex\hspace{\algorithmicindent}\Comment{4. Find the best threshold pair on the path $P$}
\State\hspace{\algorithmicindent}$l_t, r_t \gets $\text{$\underset{(l,r)\in P}{\argmin}$ \textsc{Unfiltered}$(l, r)$}
\State\hspace{\algorithmicindent}\textbf{return} $l_t, r_t$

\end{algorithmic}
\end{algorithm}

After calibrating the score distributions, we determine the optimal filtering threshold $(l, r)$, as defined in \eqref{eq:optimization}. The objective is to minimize the unfiltered rate $u(l, r)$ while satisfying the user-specified accuracy  $\alpha$. A basic brute-force search over all possible threshold pairs brings quadratic complexity. Therefore, we introduce an efficient algorithm that leverages the monotonic properties of the calibrated score distributions.
The core insight is that any optimal solution must lie on the Pareto-frontier of the feasible set defined by the accuracy constraint, $\text{Acc}(l, r) \ge \alpha$. Points off this frontier are suboptimal as their unfiltered rate can be improved by tightening the thresholds.

Our approach, outlined in Algorithm~\ref{alg:threshold-selection}, efficiently identifies the optimal thresholds by tracing the frontier. The potential thresholds, denoted as \textit{steps}, are discrete bin boundaries in our calibration. The algorithm proceeds in three main phases:

\textbf{1) Boundary Identification}: Ensuring the accuracy target, the algorithm first determines the two extreme points of the feasible frontier: $(l_0, r_s)$ and $(l_s, r_0)$. Here, $l_0$ is the tightest (highest) lower bound possible when the upper bound is most conservative ($r_s$), and $r_0$ is the lowest upper bound when the lower bound is most conservative ($l_s$). This step effectively bounds the search space to a one-dimensional path.

\textbf{2) Frontier Traversal}: Starting from $(l_0, r_s)$, the algorithm iteratively constructs a path $P$ that approximates the feasible frontier. It greedily moves towards the other extreme point $(l_s, r_0)$ by tightening either $l$ or $r$ at each step, ensuring the accuracy constraint $\alpha$ is never violated.

\textbf{3) Optimal Point Selection}: Since the unfiltered rate $u(l, r)$ is monotonically decreasing with respect to the size of the filtered region, the optimal point that minimizes $u$ must lie on the path $P$. The algorithm simply evaluates $u(l,r)$ for all points in $P$ and returns the pair $(l_t, r_t)$ that yields the minimum value.

This algorithm reduces complexity from quadratic to linear with respect to the number of discretization steps.

\one{

\newtheorem{prop}{\textbf{Proposition}}

\subsection{Theoretical Guarantee for Accuracy}
\label{s5.4}

The cascade workflow selects optimal thresholds $(l, r)$ based on a sampled subset $S'$ (with sampling ratio $p$) to satisfy a target accuracy $\alpha$. 
The thresholds are then applied back to filter the full document collection $S$. In this section, we provide a theoretical analysis of this end-to-end accuracy maintenance.

Consider a full set population $S$ of size $N$ with decision scores over $[0, 1]$. To guarantee that the true global accuracy $\text{Acc}_S$ exceeds $\alpha$ with a high confidence level ($1-\delta$), we derive a safety margin $\epsilon$ for the  estimate on $S'$.
Let $F^\pm_S(\cdot)$ and $F^\pm_{S'}(\cdot)$ denote the cumulative mass functions for the global population and the sample, respectively. $F^\pm_S$ and $F^\pm_{S'}$ represent the total mass.

The accuracy constraint (F1-score) is defined as:
\begin{equation*}
    \text{Acc}(l, r) = \frac{2(F^+ - F^+(l))}{2(F^+ - F^+(l)) + (F^- - F^-(r)) + F^+(l)} \ge \alpha
\end{equation*}
This constraint can be reformulated to bound the error distribution. We define $\mathcal{T}(l, r)$, representing the weighted sum of false positives and false negatives (the tail masses):
\begin{equation}
    \mathcal{T}(l, r) = \left(1-\frac{\alpha}{2}\right) F^+(l) + \frac{\alpha}{2} \left(F^- - F^-(r)\right) \le (1-\alpha)F^+ 
    \label{eq:tail}
\end{equation}

\begin{prop} 
Let $|S'| = pN$ be the sample size. 
For any thresholds $(l, r)$, $0\le l < r \le 1$ and $\delta>0$, there exists a $\epsilon > 0$ such that if the sample condition satisfies
$\mathcal{T}_{S'}(l, r) \le (1-\alpha)F_{S'}^+ - \epsilon$, 
then the true accuracy satisfies:
\begin{equation*}
    P[\text{Acc}_S(l, r) \ge \alpha] \ge 1 - \delta
\end{equation*}
\end{prop}

\noindent\textbf{Proof.} \hspace{0.1cm}
Let $\text{Oracle}(i) = \mathbbm{1}[i\text{-th document is positive}]$ and $s(i)$ be the score of the $i$-th document. For any pair $(l, r)$, the empirical functional $\mathcal{T}_{S'}(l, r)$ corresponds to the sample mean of i.i.d. random variables $Z_i$ over the $pN$ samples, defined as:
\begin{equation*}
    Z_i = \left(1-\frac{\alpha}{2}\right)\mathbbm{1}[\text{Oracle}(i)=1 \land s_i < l] + \frac{\alpha}{2}\mathbbm{1}[\text{Oracle}(i)=0 \land s_i > r]
\end{equation*}
By applying the Bernstein inequality, 
we can bound the deviation between the sample mean and the global mean $\mathcal{T}_S(l, r)$:
\begin{equation}
    P[|\mathcal{T}_{S'}(l, r) - \mathcal{T}_S(l,. r)| \ge \epsilon_1] \ge 1-\delta_1 \label{eq:LHS}
\end{equation}
where $\epsilon_1 = \sqrt{\frac{4\sigma_Z^2ln(2/\delta_1)}{pN}} + \frac{4(1-\alpha/2)ln(2/\delta_1)}{3pN}$, and $\sigma_Z^2=\text{Var}(Z_i)$.

Similarly, for the RHS of (\ref{eq:tail}),  we bound the estimation error of the positive class mass:
\begin{equation}
    P[|(1-\alpha)F_{S'}^+ - (1-\alpha)F_S^+|\ge \epsilon_2]\ge 1-\delta_2 \label{eq:RHS}
\end{equation}
where $\epsilon_2 = (1-\alpha) \left(\sqrt{\frac{4\sigma_P^2ln(2/\delta_2)}{pN}}+\frac{4ln(2/\delta_2)}{3pN}\right)$, $\sigma_P^2$ is the variance of the sampled positives. 

By combining (\ref{eq:LHS}) (\ref{eq:RHS}), and setting $\delta_1 = \delta_2 = \delta/2$, we establish that if the observed $\mathcal{T}_{S'}(l, r) \le (1-\alpha)F_{S'}^+ - \epsilon$, then $P[\mathcal{T}_S(l, r) \le (1-\alpha)F_S^+] \ge 1-\delta$, which implies $P[\text{Acc}_S(l, r) \ge \alpha] \ge 1-\delta$. The total margin is
$ \epsilon =
    \left(\sqrt{\sigma_Z^2} + (1-\alpha)\sqrt{\sigma_P^2} \right)\sqrt{\frac{4 \ln(4/\delta)}{pN}} + \frac{(8-6\alpha) \ln(4/\delta)}{3pN} 
    \label{eq:bound}
$.

\vspace{4pt}
\noindent\textbf{Discussion.} \hspace{0.1cm}
Proposition 1 demonstrates that the global error is bounded by the score variances, theoretically justifying our contrastive learning approach (in Section \ref{s4.1}), which optimizes for low-variance distributions. Furthermore, the discretization in Algorithm \ref{alg:threshold-selection} acts as a conservative buffer to cover the margin $\epsilon$. Collectively, this provides a theoretical guarantee that \textsc{ScaleDoc} can generalize from the proxy sample to the full dataset, maintaining the end-to-end accuracy with high confidence.
}

\section{Implementation Details}
\label{s6}

In this section, we discusses several key design decisions and optimizations within the system.

\vspace{3pt}
\noindent\textbf{Training Set Optimization.} \hspace{0.2cm} 
In the online phase, the initial step of training is to sample a small subset of documents for oracle labeling.
However, randomly sampling may get quite imbalanced classes, resulting in a biased and ineffective proxy.
To address this, we implement a fallback-style rebalancing method. 
If the initial sample is quite skewed, we augment the data by adding Gaussian noise to the existing minority embeddings and assigning them the corresponding label, thus creating a more balanced training set.
\vspace{3pt}

\noindent\textbf{Model Training.} \hspace{0.2cm} 
The proxy encoder is a 3-layer perceptron (MLP), mapping the embeddings into a latent space. Following standard practice in contrastive learning~\cite{chen2020simple, he2020momentum}, we append a projector head $Proj(\cdot)$ after the encoder, during the training. This projector head is discarded during inference.
The training mini-batch consists of multiple document embeddings and the single query embedding, forces the encoder to optimize the shared latent space for both.

In Section \ref{s4.2}, training-phase-2 optimizes a joint loss: $\mathcal{L}_{2} = \lambda\cdot\mathcal{L}_{supcon} + (1-\lambda)\cdot\mathcal{L}_{polar}$, where the hyperparameter $\lambda$ balances the two losses. Across all experiments, we empirically set $\lambda=0.2$. We found the final performance is not sensitive to this parameter, and extensive tuning did not yield significant improvements.

\vspace{3pt}
\noindent\textbf{Model Cascade.} \hspace{0.2cm} 
The calibration step relies on discretization granularity, which determines the number of bins to partition the proxy score ranges (the $steps$ in Algorithm \autoref{alg:threshold-selection}). We empirically set this granularity to 64,  balancing distribution accuracy with sample representativeness per bin.


\section{Evaluation}
\label{s7}


\begin{figure*}[t!]
  \centering
  \includegraphics[width=0.88\linewidth]{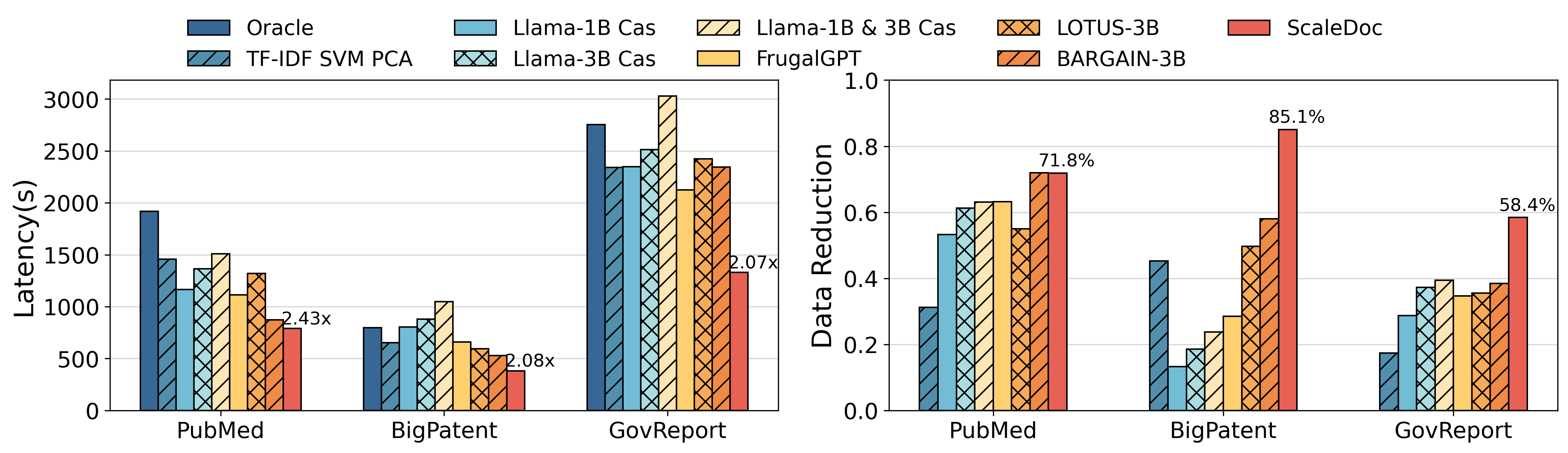}
  \caption{\two{End-to-end latencies and data reduction rate -- \textmd{We evaluate \textsc{ScaleDoc} and other baselines with accuracy target $\alpha$ = 0.90. The data reduction rate measures the percentage of data that does not require the LLM oracle call, indicating the \textbf{cost-saving}. }}}
  \label{fig:e2e}
\end{figure*}
\subsection{Experimental Setup}
\label{s7.1}

\noindent\textbf{Workloads.}
\begin{table*}[t!]
  \caption{Dataset Characteristics}
  \label{tab:ds}
  \begin{tabular}{l p{5cm} c p{7cm}}
    \toprule
    Dataset & Content & Avg. Word Count & Example Predicating Query \\
    \midrule
    \textsc{PubMed} &  Abstracts from medical papers  & 413 & Report efficacy of a certain medicine?
 \\
    \textsc{BigPatent} & Summaries on U.S. patent documents & 129 & Introduce inventions of cross-sectional technology? \\
    \textsc{GovReport} & Reports from government  & 621 & About environment protection? \\
    \bottomrule
  \end{tabular}
\end{table*}
 To evaluate performance across diverse settings, we use three real-world document collections: \textsc{BigPatent}~\cite{sharma-etal-2019-bigpatent}, \textsc{PubMed} \cite{dernoncourt-lee-2017-pubmed}, and \textsc{GovReport}~\cite{huang-etal-2021-efficient}. Given the lack of benchmarks for collection-level semantic predicates,  we manually crafted 20 distinct natural language queries for each collection, covering a wide range of semantic characteristics and data selectivity.
Each collection contains 10,000 documents, and the ground truth labels for all query-document predicating pairs are generated by GPT-4o ~\cite{openai2024gpt4ocard}.
\autoref{tab:ds} provides a summary of the datasets and illustrative queries. We utilize \textbf{F1 score as the accuracy metric} throughout our paper, which is more robust than naive accuracy for handling imbalanced data.

\vspace{4pt}
\noindent\textbf{Baselines.}
We compare \textsc{ScaleDoc} against several representative baselines, which accelerate  ML-powered predicates in analytical data systems. The oracle model for all systems, which also serves as the ground truth labeler, is \textbf{GPT-4o}~\cite{openai2024gpt4ocard}.

\textbf{1) Oracle Only.} \hspace{0.1cm}
This baseline avoids any filtering and processes every document directly using the oracle LLM. We prompt the LLM to output the binary class labels to indicate the predicate decision.

\textbf{2) Probabilistic Predicates (PPs)} ~\cite{lu2018accelerating, yang2022optimizing}.\hspace{0.1cm}
This approach uses traditional lightweight machine learning models as proxies. For each new problem, the PPs are trained to  output binary classification results with confidence interface.
It includes the following component choices:
\begin{itemize}[leftmargin=1.5em,itemsep=1pt,topsep=1pt]
\item Text Representation: Bag-of-Words (BoW) and TF-IDF, both are standard text encoding schemes.
\item Dimensionality Reduction: Principal Component Analysis (PCA) and Feature Hashing (FH).
\item Classifier: Support Vector Machines (SVM) and Kernel Density Estimation (KDE).
\end{itemize}
These models are implemented using scikit-learn ~\cite{pedregosa2011scikit}. We explore various combinations of them and report the best performing choices.

\textbf{3) LLM Cascade}~\cite{chen2023frugalgptuselargelanguage, 10.14778/3749646.3749685}. \hspace{0.1cm}
This category employs smaller, cost-effective LLMs as proxies to filter easy instances, selectively deferring only uncertain cases to the oracle.
\two{In our setup, Llama-3.2-1B or 3B models serve as proxies, while GPT-4o acts as the oracle.  We utilize the \textsc{vLLM} engine~\cite{kwon2023efficient} to ensure high-throughput proxy inference.
We evaluate four distinct cascade strategies:}

\begin{itemize}[leftmargin=1.5em,itemsep=2pt,topsep=2pt]
    \item \two{\textbf{\textsc{Basic Cascade.}}  We implement standard cascade pipelines, including single-proxy configurations (denoted as \textit{1B-Cas} and \textit{3B-Cas}) and a multi-hop chain (\textit{1B$\to$3B$\to$Oracle}). These baselines utilize the log-probability of the generated token as the confidence score. To determine the filtering thresholds, they adopt the same cascade strategy described in Section 4.}
    \item \two{\textbf{\textsc{FrugalGPT}}~\cite{chen2023frugalgptuselargelanguage}. This approach employs a scoring model (distilBERT ~\cite{sanh2019distilbert}) to predict the reliability of the proxy LLM's response. It then uses a constrained optimization formulation to maximize accuracy within a given budget.
    We integrate both 1B and 3B models as proxies, and \textsc{FrugalGPT} would autonomously build a cascading pipeline. To ensure a fair comparison with our accuracy-target setting, we profile the cost-accuracy curve of \textsc{FrugalGPT} and report the minimum oracle usage required to achieve the target accuracy.}
    \item \two{\textbf{\textsc{LOTUS}}~\cite{10.14778/3749646.3749685}.  With a database perspective, \textsc{LOTUS} implements semantic operators and adopts a variant of the SUPG algorithm~\cite{10.14778/3407790.3407804} to accelerate predicates. Similar to \textsc{ScaleDoc}, it relies on an independent sampling procedure to estimate statistics and derive cascade thresholds over the sampled subset.}
    \item \three{\textbf{\textsc{BARGAIN}}~\cite{10.1145/3769776}.
    This recent framework provides theoretical statistical guarantees for cascade flows. }
    \M{It introduces different optimization targets (i.e., precision, recall and accuracy target). To align with our setting, we compare against its accuracy-target (AT) strategy, which minimizes oracle usage while satisfying a correctness constraint. Since its AT strategy optimizes for exact-match accuracy (\textsc{ScaleDoc} use F1 score by default), we implement an exact-match variant of \textsc{ScaleDoc} to align the accuracy metrics and report the normalized results.}
\end{itemize}

\textbf{4) Direct Embedding Matching}. \hspace{0.1cm}
For additional ablation, we adopt a design from embedding-based retrieval systems \cite{huang2020embedding-ret}. This method computes a similarity score between the query and each document using an \textit{off-the-shelf} embedding model. This score serves directly as the proxy value for cascade filtering. (See results in Section \ref{s7.4}.)

\vspace{4pt}
\noindent\textbf{Experimental Details.}
Each workload operates on a collection of 10,000 documents. For methods requiring proxy fine-tuning (i.e., \textsc{ScaleDoc} and PPs), we sample 1,000 documents (10\%) for training and calibration. The remaining 9,000 documents are then processed online.
Baselines that do not require training (i.e., oracle-only and LLM-cascade) just process the entire collection online.
For the offline representation phase, \textsc{ScaleDoc} uses \textsc{NvEmbed}~\cite{lee2024nv}, derived from Mistral-7B~\cite{jiang2023mistral7b}, to pre-compute the document embeddings. Our online lightweight proxy model is implemented as a 3-layer MLP.
We run LLM inference on a single NVIDIA A10 GPU, while the oracle LLM (GPT-4o) is accessed via the Azure-OpenAI API, subject to a rate limit of 150k tokens per minute.

\subsection{End-to-End Performance}
\label{s7.2}


We evaluate end-to-end online performance of \textsc{ScaleDoc} against the baselines. Figure \ref{fig:e2e} presents the average online latency and data reduction across three datasets, with a user-specified accuracy target of $\alpha=0.90$ ($\alpha$ for all experiments will be 0.90 if not specified). The results show that \textsc{ScaleDoc} consistently and significantly outperforms all other approaches, achieving an average online \textbf{speedup of over 2$\times$}. \textsc{ScaleDoc} also reduces the invocation of oracle LLMs up to \textbf{85\%}, which means \textbf{6.6$\times$ cost-saving}.
\M{Specifically, \textsc{ScaleDoc} achieves better oracle reduction in 44 out of 60 queries  against BARGAIN-3B and outperforms LOTUS-3B in 53 out of 60 queries.}
The performance gain is driven by effective data filtering  and the computational efficiency of our lightweight design.

\begin{table}[t!]
\caption{
Estimated average computational cost (FLOPs) per query - \textmd{The FLOPs are normalized to 10,000 documents. Each column present the overall averaged invocation frequency of the model. ``Total'' represents the sum of all components. (The labeling cost is omitted.)}}
\label{tab:flops}
\begin{tabular}{c | c | c | c | c | c}
\toprule
\begin{tabular}{@{}c@{}}Models($\to$) \\ FLOPs \end{tabular} & \begin{tabular}{@{}c@{}}Our Proxy \\ 2T\end{tabular} & \begin{tabular}{@{}c@{}}1B \\ 10P\end{tabular} & \begin{tabular}{@{}c@{}}3B \\ 27P\end{tabular} & \begin{tabular}{@{}c@{}}Oracle \\ >500P\end{tabular} & Total \\
\midrule
\textbf{ScaleDoc} & \textbf{1$\times$} & \textbf{-} & \textbf{-} & \textbf{0.28$\times$} & \textbf{140P} \\
1B \& 3B Cas & - & 1$\times$ & 0.42$\times$ & 0.59$\times$ & 316P\\
LOTUS-3B & - & - & 1$\times$ & 0.61$\times$ & 332P \\
BARGAIN-3B & - & - & 1$\times$ & 0.44$\times$ & 249P \\
Oracle & - & - & - & 1$\times$ & 500P\\
\bottomrule
\end{tabular}
\end{table}

\vspace{3 pt}

\noindent\one{\textbf{Computational Efficiency.}\hspace{0.2cm}
For a more holistic evaluation of \textsc{ScaleDoc}'s cost-effectiveness,  beyond latency and data reduction ratios, we analyze the total computation consumption. We estimate the Floating Point Operations (FLOPs) required per query, with the averaged document length as input (i.e., approximately 400 words). Table~\ref{tab:flops} details the invocation frequency of different models and the resulting total FLOPs (normalized to 10,000 documents). The results demonstrate that \textsc{ScaleDoc} achieves the lowest total computational cost (140P), significantly outperforming other strategies.
This efficiency stems from the synergy of two factors: the negligible training and inference cost of our tiny proxy model; and the substantial reduction in expensive Oracle LLM invocations.
}

\begin{figure}[t!]
  \centering
    \includegraphics[width=\linewidth]{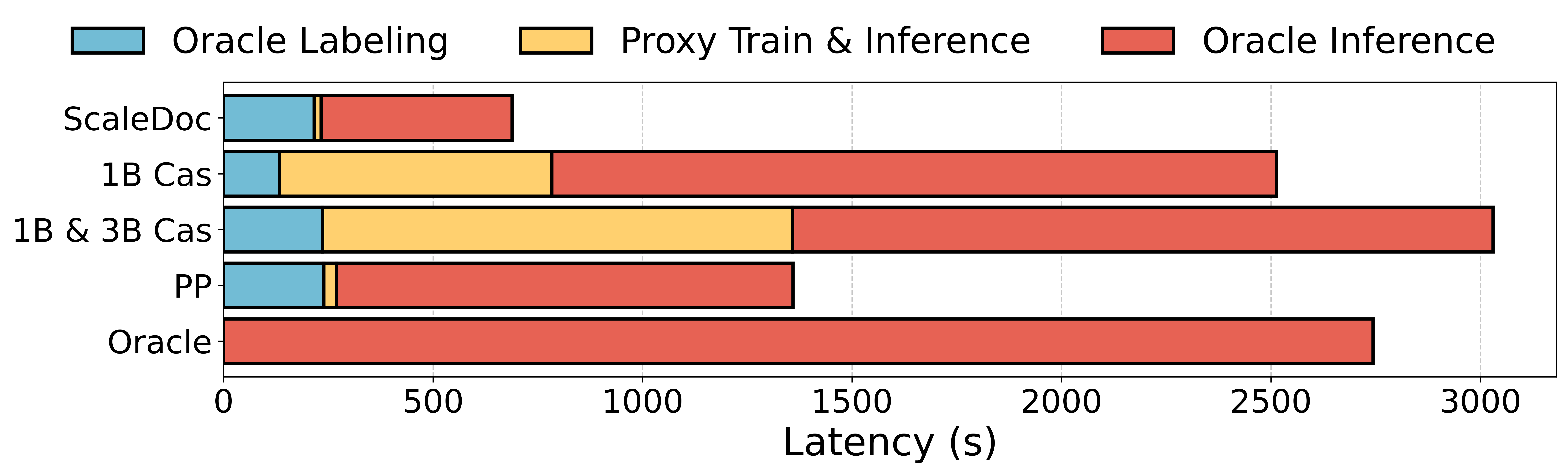}
  \caption{
  Breakdown for different approaches over \textsc{PubMed} dataset, measuring average latencies of each stage. -- \textmd{{\textsc{ScaleDoc} (top)} presents significance improvement.}
  }
  \label{fig:e2e-detail}
\end{figure}

\vspace{3 pt}
\noindent\textbf{Latency Breakdown.}\hspace{0.2cm} \three{
To provide a granular performance analysis, Figure \ref{fig:e2e-detail} breaks down the average latency of each online processing stage over \textsc{PubMed} datasets. We report 3 different stages of the pipeline, including oracle labeling, proxy train \& inference and oracle inference. The oracle-labeling stage represents the overhead of invoking the oracle to label a sample subset for subsequent proxy training and calibration.}
Compared to other approaches, \textsc{ScaleDoc} achieves significant {reduction in oracle inference}. Its efficiency is also driven by its {lightweight proxy model}, which benefits training and inference.
In contrast, the LLM cascade methods exhibit a  performance bottleneck. Although their proxy LLMs offer training-free zero-shot capabilities, they still incur moderate computational costs during inference, leading to high end-to-end latency.

\vspace{4 pt}
\noindent\textbf{Offline Overhead.}\hspace{0.2cm}
\textsc{ScaleDoc} needs an offline computation for document representation. Encoding 10,000 documents with \textsc{NvEmbed} is a \textbf{one-time} process and requires a relatively small amount of computation.
For example, on the \textsc{PubMed} dataset, the estimated computation  is approximately 50 PFLOPs.
 Conversely, during online processing, an LLM oracle (e.g., GPT-4o) demands 10$\times$ more computation for each ad hoc query.
 A lightweight proxy model such as Llama-3.2-3B would incur 27 PFLOPS online computation for each new query.
 This comparison underscores that \textsc{ScaleDoc}'s one-time offline overhead is relatively small and acceptable, in light of the efficiency it brings to online query processing.

\subsection{\three{Accuracy Analysis}}
\label{s7.3}

\begin{figure}[t!]
  \centering
  \begin{subfigure}{0.49\linewidth}
  \subfloat{
    \includegraphics[width=\linewidth]{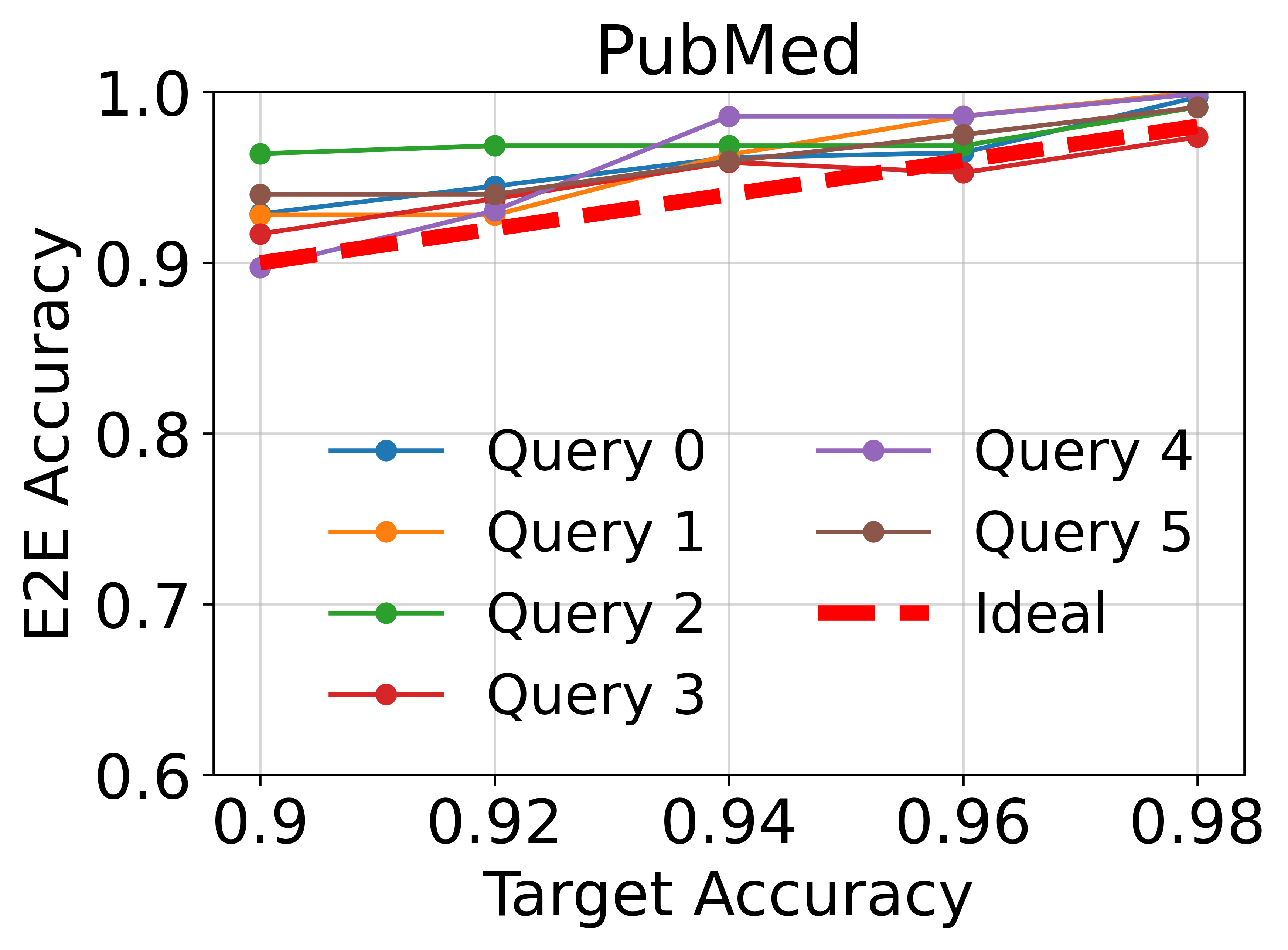}
    }
  \end{subfigure}
  \hfill
  \begin{subfigure}{0.49\linewidth}
  \subfloat{
    \includegraphics[width=\linewidth]{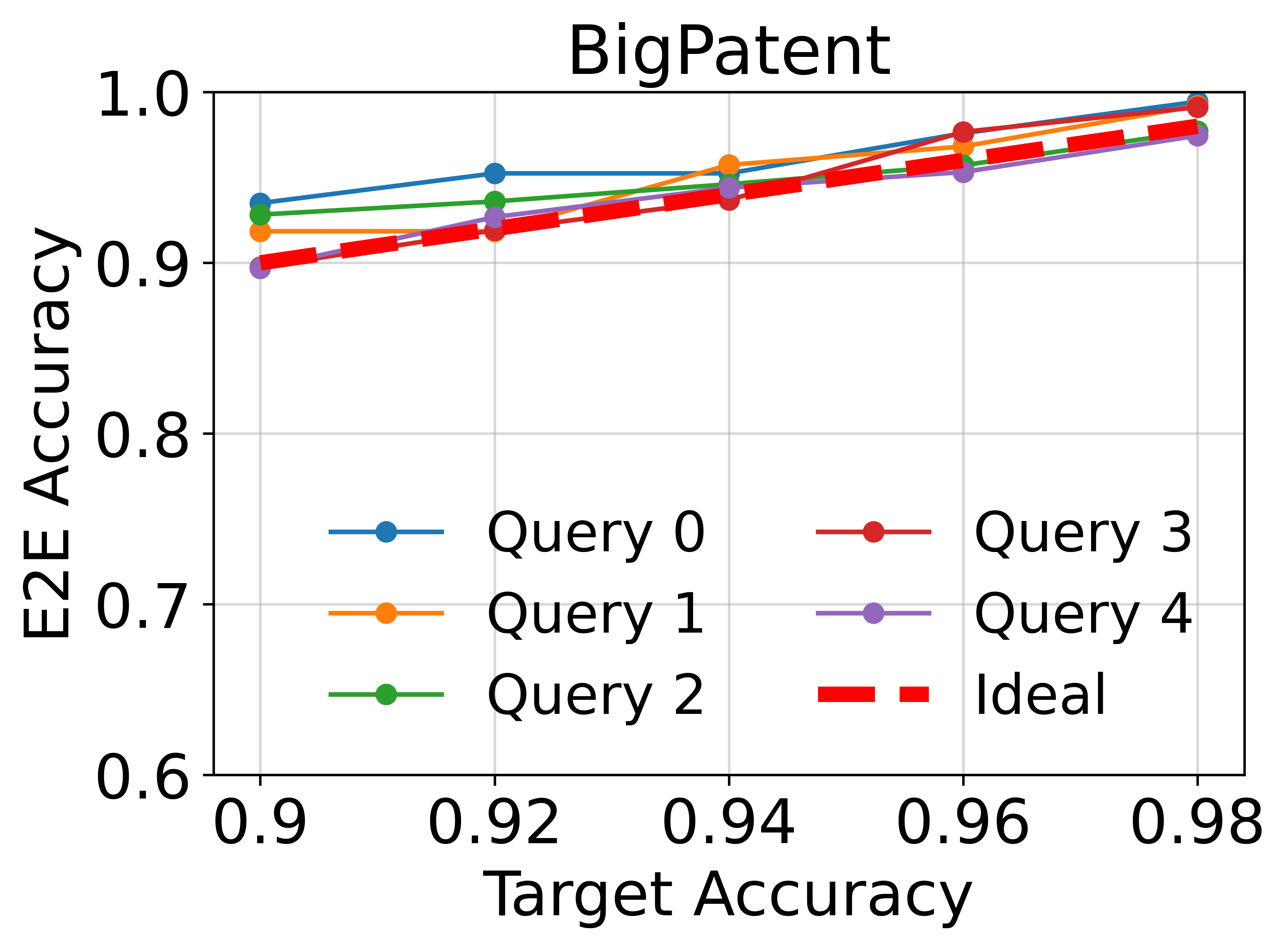}
    }
  \end{subfigure}
  \caption{\three{Accuracy Validation with human-annotated labels.}}
  \vspace{-15pt}
  \label{fig:gt}
\end{figure}

\three{
\noindent\textbf{Validation against Human Labels.}\hspace{0.2cm}
Since human-labeled data is typically limited to specific classification tasks, our main evaluation relies on various manually curated queries.
 Nevertheless, to ensure rigorous validation against human standards, we conducted additional experiments using human annotations from the \textsc{BigPatent} and \textsc{PubMed} datasets.
We mapped the original classification labels to binary predicates and evaluated the actual end-to-end accuracy of \textsc{ScaleDoc}.
\autoref{fig:gt} demonstrates that our system successfully maintains user-specified accuracy (achieving the \textit{Ideal} dotted line).
Besides, across these 11 queries with a target accuracy of $\alpha=0.9$, \textsc{ScaleDoc} reduces average oracle invocations by 80.6\%.
These results confirm that our approach is robust against human judgment and maintains high efficiency.}

\begin{figure}[t!]
  \centering
    \includegraphics[width=0.85\linewidth]{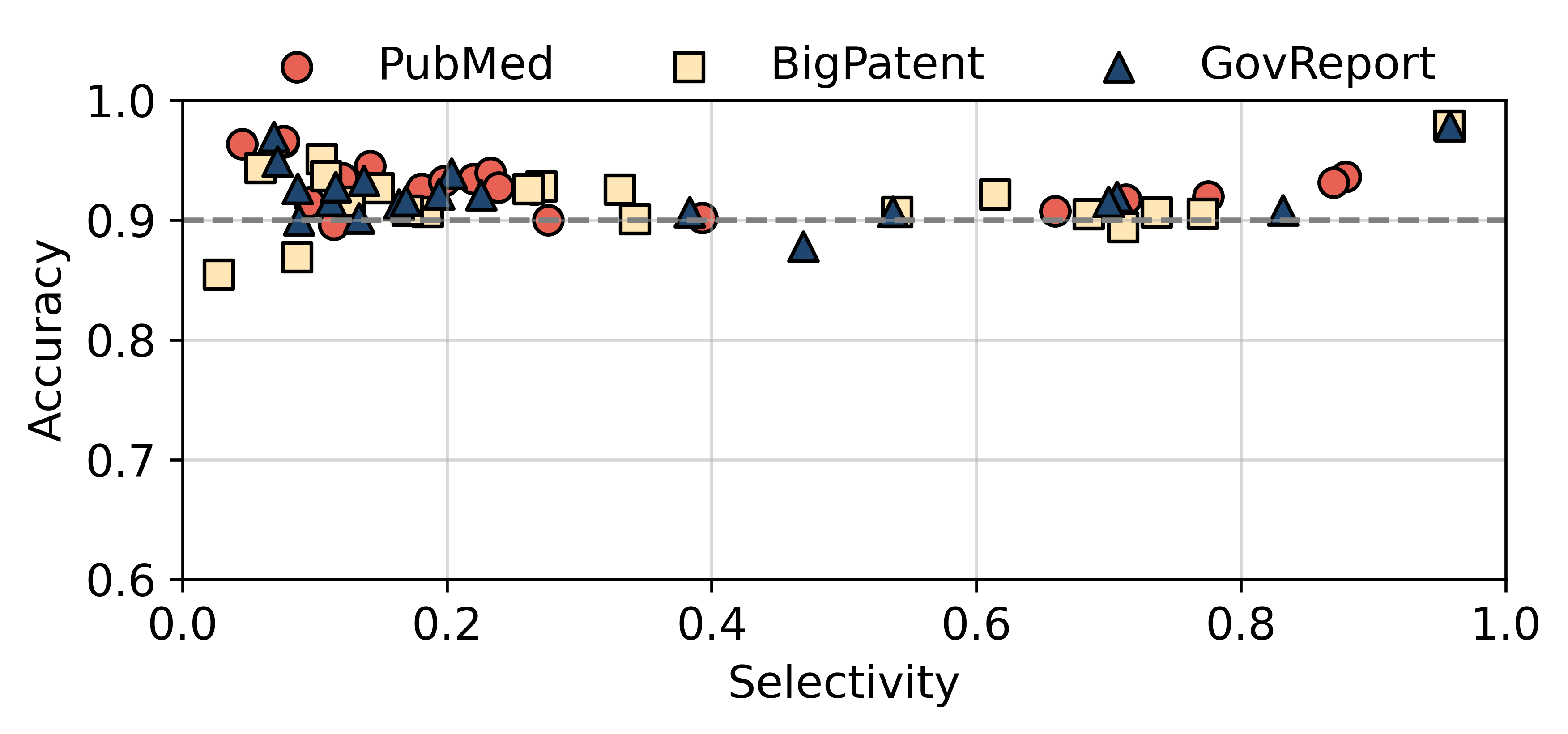}
    \caption{\three{Accuracy performance over different selectivities.}}
    \label{fig:acc-selec}
\end{figure}

\noindent \three{\textbf{Accuracy Robustness across Selectivities.}\hspace{0.3cm}
For predicate evaluation, selectivity denotes the fraction of positive instances, thereby introducing challenges related to class imbalance. To assess robustness against such skewness, Figure \ref{fig:acc-selec} reports the accuracy of \textsc{ScaleDoc} across varying selectivities. The results demonstrate that \textsc{ScaleDoc} exhibits resilience to data skew, maintaining robust accuracy across a broad range of selectivities. (More details about skewness are presented in Section~\ref{s7.6}.)}

\begin{figure}[t!]
  \centering
  \includegraphics[trim = 0.1cm 3cm 0.1cm 3.5cm , clip, width=0.98\linewidth]{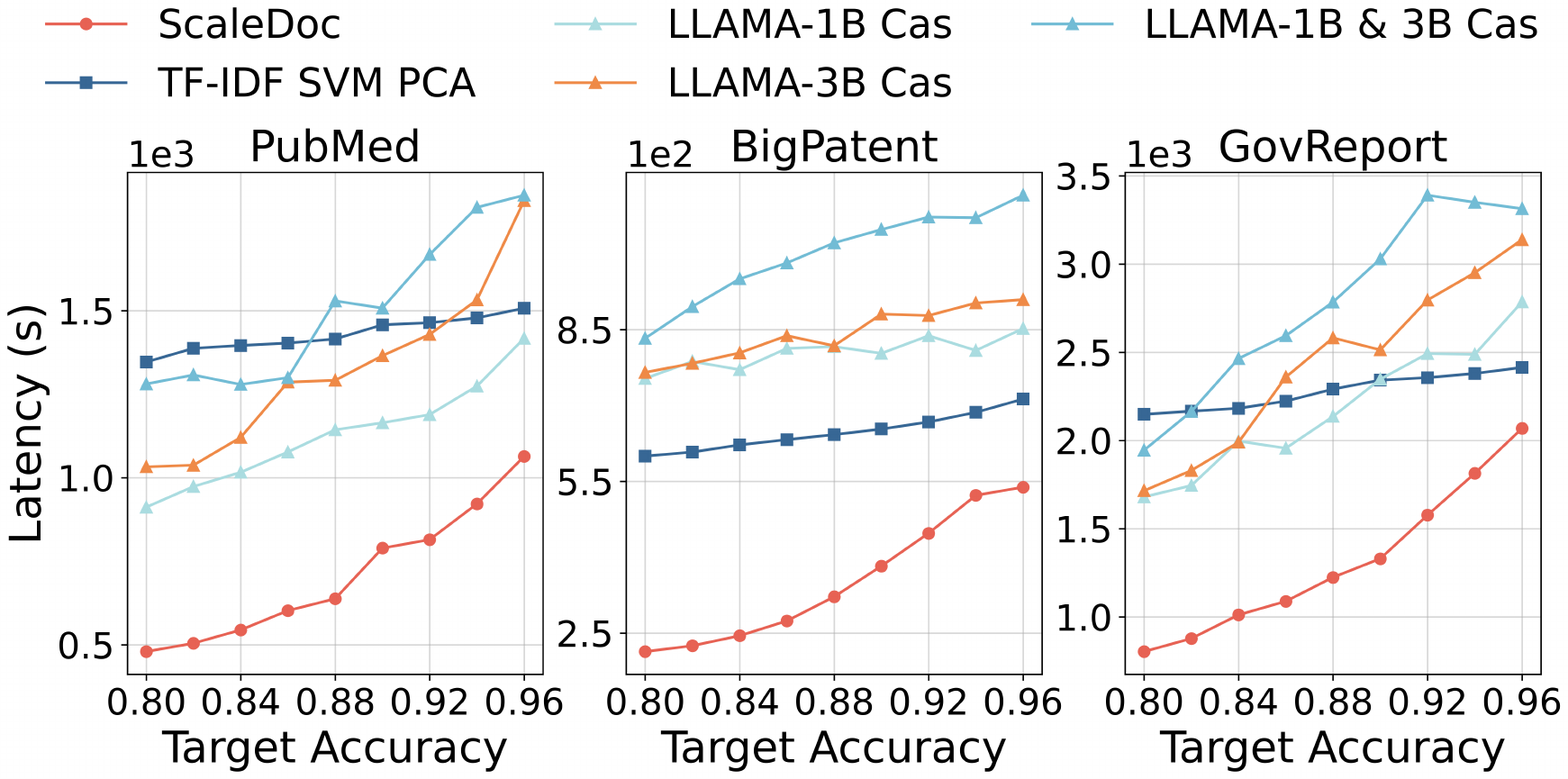}
  \caption{Accuracy-Latency tradeoff on three datasets.}
  \label{fig:tradeoff}
\end{figure}

\vspace{3pt}
\noindent\textbf{Accuracy-Latency Tradeoff.}\hspace{0.3cm}
We evaluated the accuracy-latency trade-off by setting accuracy targets from 0.80 to 0.96. As shown in \autoref{fig:tradeoff}, \textsc{ScaleDoc} consistently reduces average runtime across three datasets and different accuracy targets.
All of the methods exhibit a common trend: higher accuracy targets generally incur more latency. When accuracy targets are relaxed, \textsc{ScaleDoc}'s latency decreases by a larger margin, offering more trade-off opportunities for higher performance. In contrast, other baselines show smaller latency variation, due to their unreliable decision scores. This limits their data reduction even when accuracy requirements are loosened.

\subsection{Ablation for Proxy Model Training}
\label{s7.4}

\vspace{4 pt}
\noindent\textbf{Ablation Comparison.}\hspace{0.2cm}
To assess the impact of our overall training framework, we compare it against the baselines that rely on direct embedding-based similarity matching. Specifically, we evaluate on \textsc{NvEmbed}, the embedding model used in \textsc{ScaleDoc}'s offline phase, along with a widely adopted model, E5~\cite{wang2024multilinguale5textembeddings, wang2024textembeddingsweaklysupervisedcontrastive}. For this approach, similarity scores are directly computed between each document and the query, then serving as proxy values for cascading decisions.
As shown in \autoref{tab:embed_sim}, our query-aware training  outperforms these direct approaches. Unlike static latent representations, our paradigm dynamically adapts to online query semantics, producing more reliable decision scores and superior efficiency.


\begin{table}[t!]
\caption{End-to-end processing latencies of two direct embedding-matching methods and \textsc{ScaleDoc}.}
\label{tab:embed_sim}
\resizebox{0.7\linewidth}{!}{
\begin{tabular}{c | c | c | c}
\toprule
 & PubMed & BigPatent & GovReport \\
\midrule
E5    & 1412.0 & 708.1 & 2257.6 \\
NvEmbed & 1358.9 & 641.2 & 2206.5 \\
\textbf{ScaleDoc} & \textbf{789.1} & \textbf{382.6} & \textbf{1330.6} \\
\bottomrule
\end{tabular}
}
\end{table}

\begin{figure}[t!]
  \centering
  \includegraphics[trim = 2.5cm 3.5cm 2.5cm 3.5cm , clip, width=0.9\linewidth]{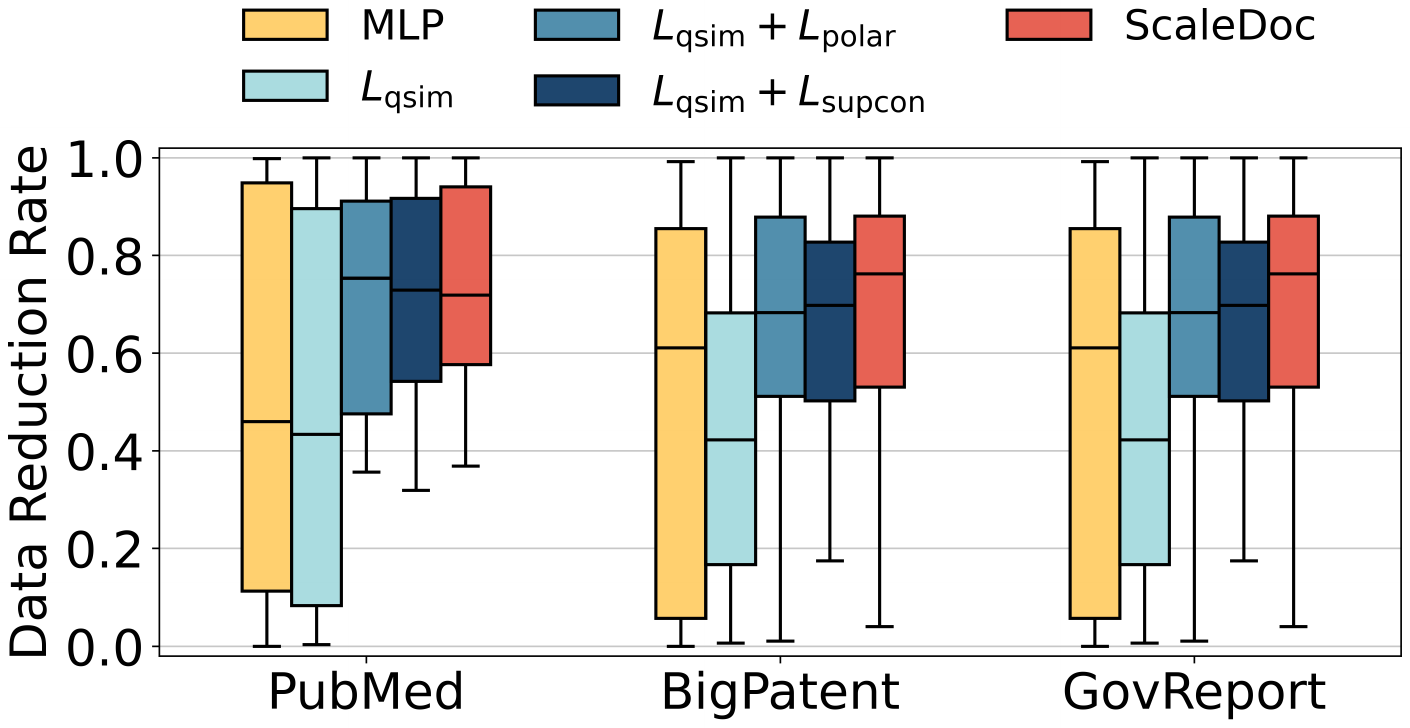}
  \caption{
  End-to-end data reduction ablation results on different training loss variants -- \textmd{\textsc{MLP} serves as a baseline, training a standard binary classifier. $\mathcal{L}_{qsim}$, $\mathcal{L}_{supcon}$, $\mathcal{L}_{polar}$ are ablation-style variants of \textsc{ScaleDoc}'s training approach.}}
  \label{fig:cldr}
\end{figure}

\vspace{4 pt}
\noindent\textbf{Breakdown of the Training.}\hspace{0.3cm}
We further provide a decomposed analysis of the proposed contrastive learning objectives.
\autoref{fig:cldr} shows the effectiveness of our training design through end-to-end data reduction results. Here, we use a brute-force optimal cascade to isolate the effect from cascade designs. The key findings include:

\textbf{1) Superiority of contrastive-based paradigm.}
We compare \textsc{ScaleDoc} with a baseline where a MLP binary classifier is directly trained on document embeddings.
 \autoref{fig:cldr} shows that \textsc{ScaleDoc} achieves 20\% higher data reduction than basic MLP.
Our further investigation suggests that the standard MLP training is inadequate for handling the complex query-document fusion task.

\textbf{2) Effectiveness of different losses.} Our ablation (\autoref{fig:cldr}) reveals the distinct roles of each loss. Training with only $\mathcal{L}_{qsim}$ in phase 1 is insufficient, but phase 2, with $\mathcal{L}_{supcon}$ and $\mathcal{L}_{polar}$, resolves this to achieve the final performance. This does not diminish the role of phase 1. In fact, models trained without phase 1 failed to produce valid results, demonstrating that establishing an initial semantic ranking is a fundamental prerequisite.

\begin{figure}[t!]
  \centering
    \includegraphics[width=\linewidth]{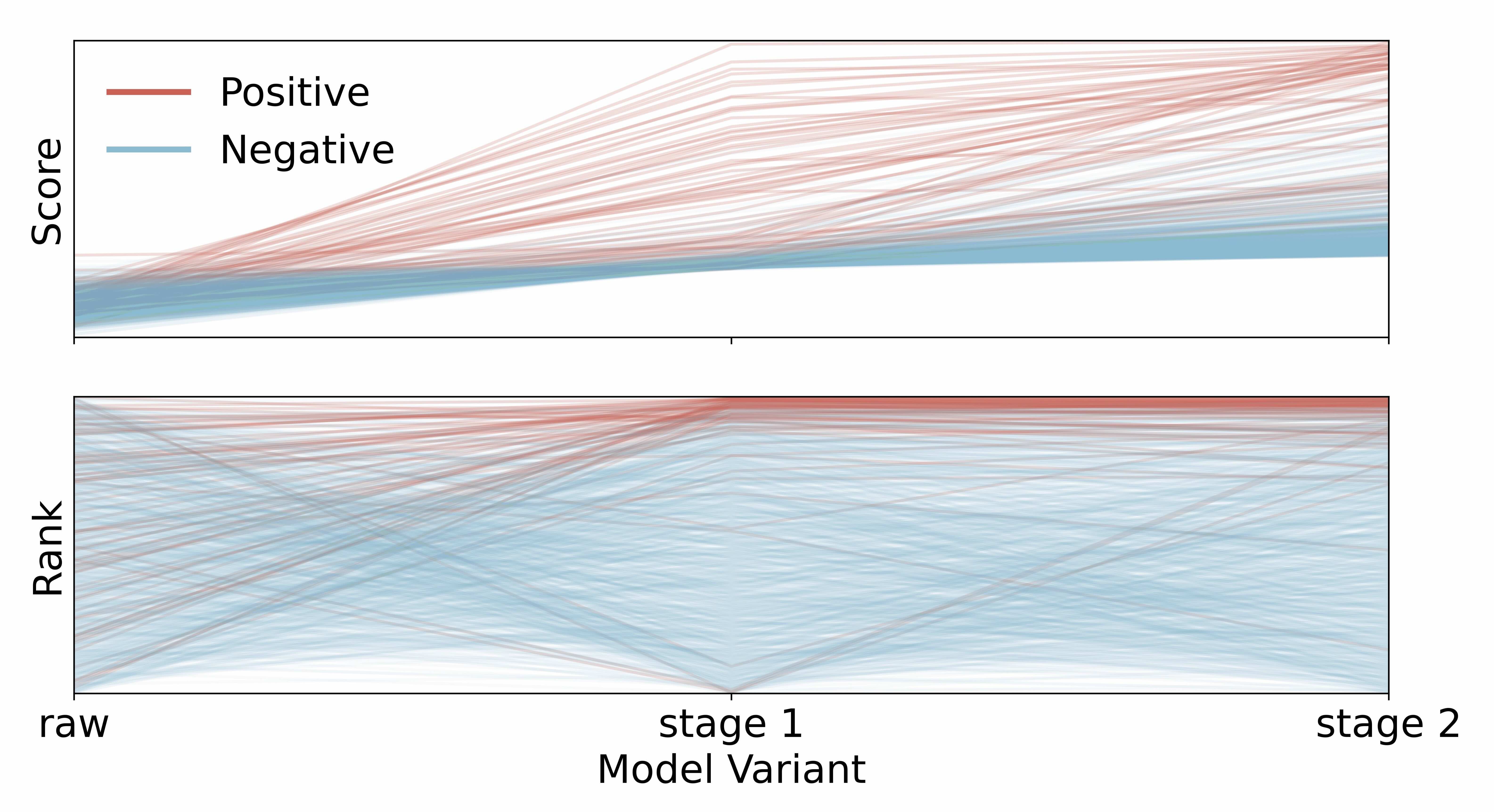}
  \caption{Embedding relocation mapping during training. -- \textmd{\textsc{Score} stands for the numerical values of cosine similarity scores, and \textsc{Rank} stands for the relative order of the scores. Each line demonstrates the relocation trace of a single document in the latent space throughout 2 training phases.}}
  \label{fig:mono}
\end{figure}

We dive deeper into these objectives with following experiments:


\textbf{a) $ \mathcal{L}_{qsim}$ guarantees semantic monotonicity.} \hspace{0.1cm}
\autoref{fig:mono} illustrates how the lightweight encoder refines document representations. The encoder maps the original embedding (\textsc{raw}) to a latent space where scores reflect semantic consistency to the predicates.
While positive and negative documents initially overlap, stage-1 training with $\mathcal{L}_{qsim}$ makes them semantically orderly: positive documents shift to high-score regions, and negatives to low ones. This establishes \textit{semantic monotonicity}, forming a crucial foundation.


\begin{figure}[t!]
  \centering
    \includegraphics[trim=0.1cm 4.5cm 0.1cm 4.5cm, clip, width=0.7\linewidth]{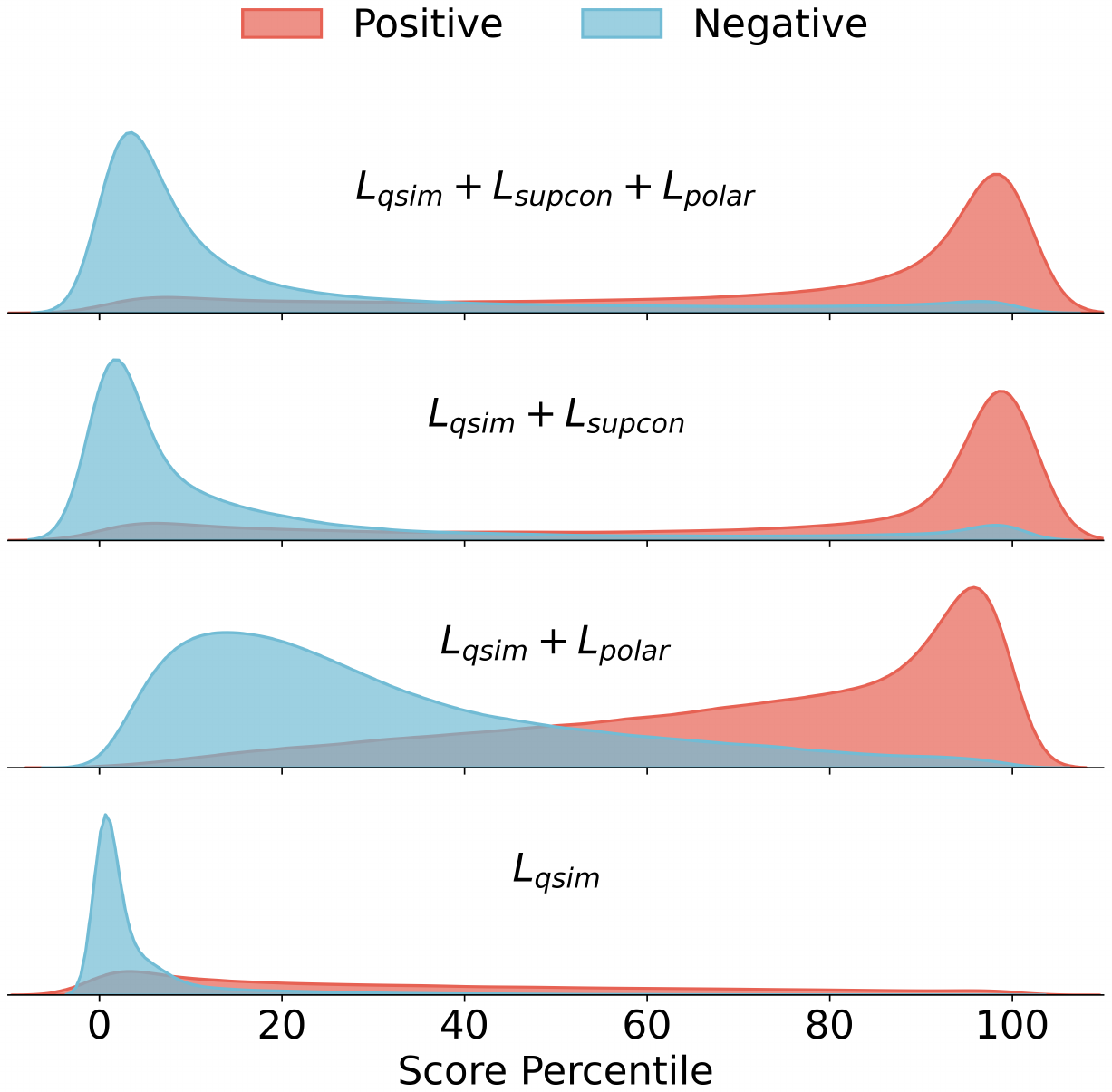}
  \caption{Average score distribution of positives and negatives -- \textmd{The four model variants are trained with different losses in an ablation style. All scores are normalized into percentiles.}}
  \label{fig:abl1}
\end{figure}

\textbf{b) $\mathcal{L}_{supcon}$ and $\mathcal{L}_{polar}$ create a bipolar distribution.} \hspace{0.1cm}
\autoref{fig:abl1} illustrates score distributions under different training objectives. $\mathcal{L}_{qsim}$ alone yields overlapping positive and negative distributions, complicating threshold selection. Both $\mathcal{L}_{supcon}$ and $\mathcal{L}_{polar}$ address this by creating a \textit{bipolar distribution},
pushing positive and negative scores to opposite ends. Specifically, $\mathcal{L}_{supcon}$ clusters documents with the same label and creates \textit{high-kurtosis peaks},  but it introduces small and incorrect sub-clusters in the tail regions. $\mathcal{L}_{polar}$, in contrast, does not cluster as aggressively, but is  effective at shaping the tails. This creates a cleaner separation and mitigates the side effects of $\mathcal{L}_{supcon}$. Integrating all three losses together achieves the reliable final distribution


\subsection{Validation of the Cascade Mechanism}
\label{s7.5}

\begin{figure}[t!]
  \centering
  \begin{subfigure}{\linewidth}
    \includegraphics[ width=\linewidth]{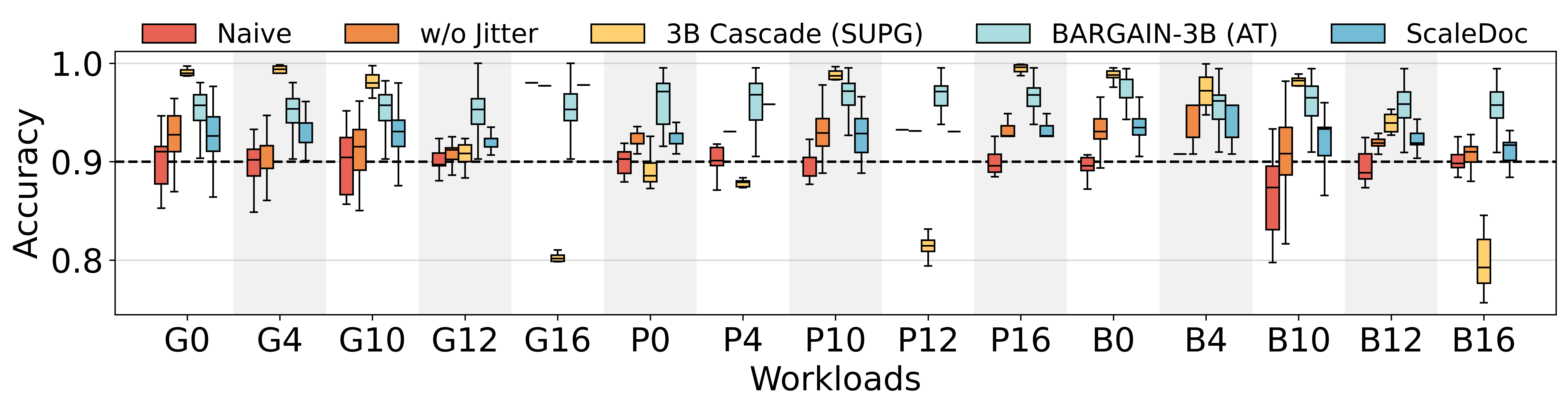}
    \caption{\one{Cascade accuracy on 100 trials -- \textmd{Black dotted line demonstrate accuracy target $\alpha$=0.90.} }}
    \label{fig:zs1}
  \end{subfigure}
  \hfill
  \begin{subfigure}{\linewidth}
  \centering
    \includegraphics[trim = 0cm 0.5cm 0cm 1cm , clip, width=0.85\linewidth]{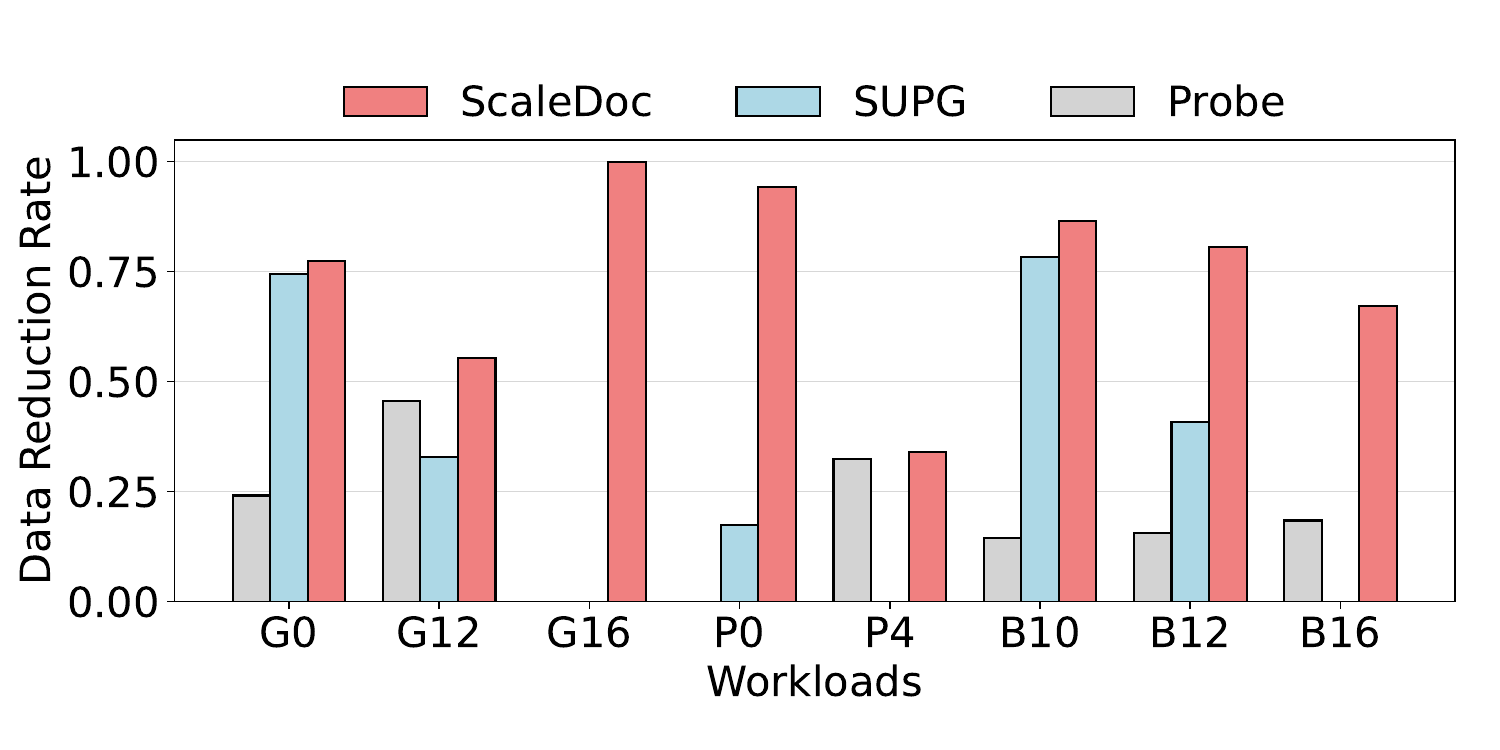}
    \caption{Data reduction over different cascade algorithms -- \textmd{\textsc{SUPG} and \textsc{Probe} may result in zero data reduction, represented by blank bars.} }
    \label{fig:zs2}
  \end{subfigure}
  \caption{Ad hoc cascade accuracy and data reduction rates.}
  \label{fig:zs-valid}
\end{figure}

\noindent
\textbf{Accuracy Maintenance.} \hspace{0.2cm}
\one{
Handling ad hoc cascade is inherently challenging because the system lacks prior knowledge to guarantee specified accuracy targets.
To empirically validate the reliability of our cascade accuracy, we compare it against SUPG~\cite{10.14778/3407790.3407804}, BARGAIN~\cite{10.1145/3769776}, a \textit{Naive} approach (directly selecting thresholds from the sampled distribution), and an ablation variant of our approach (\textit{w/o Jitter}).
We evaluate 100 trials over 15 queries, reporting F1 scores, whereas BARGAIN uses the accuracy ratio.}
\one{As shown in \autoref{fig:zs1}, the \textit{Naive} approach fails to achieve the target accuracy in nearly half of the trials due to uncontrolled calibration.
Similarly, both the \textit{w/o Jitter} variant and SUPG exhibit instability across several queries.
Although SUPG leverages statistical bounds for threshold selection, it misses the target in some practical cases due to information bias.
Critically, Both BARGAIN and our \textsc{ScaleDoc} maintain a robust accuracy to achieve the specified target. These results confirm  \textsc{ScaleDoc}'s reliability for accuracy maintenance.
}

\vspace{3pt}
\noindent\textbf{Effect on Data Reduction.} \hspace{0.2cm}
Furthermore, we evaluate the effectiveness of different cascade mechanisms in reducing LLM invocation, given the same embedding-based proxy.
We introduce an additional baseline, \textit{Probe-based Calibration}, which iteratively forwards the most ambiguous documents to the oracle (starting from a decision score of 0.5).
\autoref{fig:zs2} demonstrates that \textsc{ScaleDoc} delivers superior and generalizable performance.
The SUPG and Probing methods struggle with some cases, even yielding zero data reduction.
This further validates the effectiveness of our adaptive cascade strategy in maximizing efficiency under accuracy constraints.

\vspace{3 pt}
\noindent\textbf{Validate Density Estimator.} \hspace{0.2cm}
The effectiveness of ad hoc calibration is based on a faithful Density Estimator (DE) for the score distribution. A good DE should accurately model original distributions from a small subset.
To validate its effectiveness, we leverage Jenson-Shannon Distance (JSD) to measure the discrepancy between the reconstructed distribution and the original full distribution. We compare the quality of \textsc{ScaleDoc}'s DE results (\textsc{SD}) with other estimators, such as Importance Sampling (\textsc{IS)} and directly fitting Beta distribution curves (\textsc{B}). As shown in \autoref{tab:recon}, across 10 different queries, our DE (optimized with linear interpolation) achieves the lowest discrepancy in general, ensuring the selected thresholds are accurately transferable to the full set.

\begin{table}[t!]
\caption{JSD between reconstructed distribution and ground truth (Lower is better). -- \textmd{\textsc{N} denotes Naive stratified sampling. p/n denotes positive and negative distributions.}}
\label{tab:recon}
\resizebox{0.95\linewidth}{!}{
\begin{tabular}{c | c c | c c | c c | c c}
\toprule
 & N-p & N-n & IS-p & IS-n & B-p & B-n & SD-p & SD-n \\
\midrule
Mean    & 0.30 & 0.11 & 0.34 & 0.18 & 0.50 & 0.38 & \textbf{0.20} & \textbf{0.09} \\
Median & 0.32 & 0.08 & 0.33 & 0.17 & 0.42 & 0.43 & \textbf{0.16} & \textbf{0.08} \\
\bottomrule
\end{tabular}
}
\end{table}

\subsection{Latency Analysis of the Data Skewness}
\label{s7.6}
\begin{figure}[t!]
  \centering
    \includegraphics[width= 0.9\linewidth]{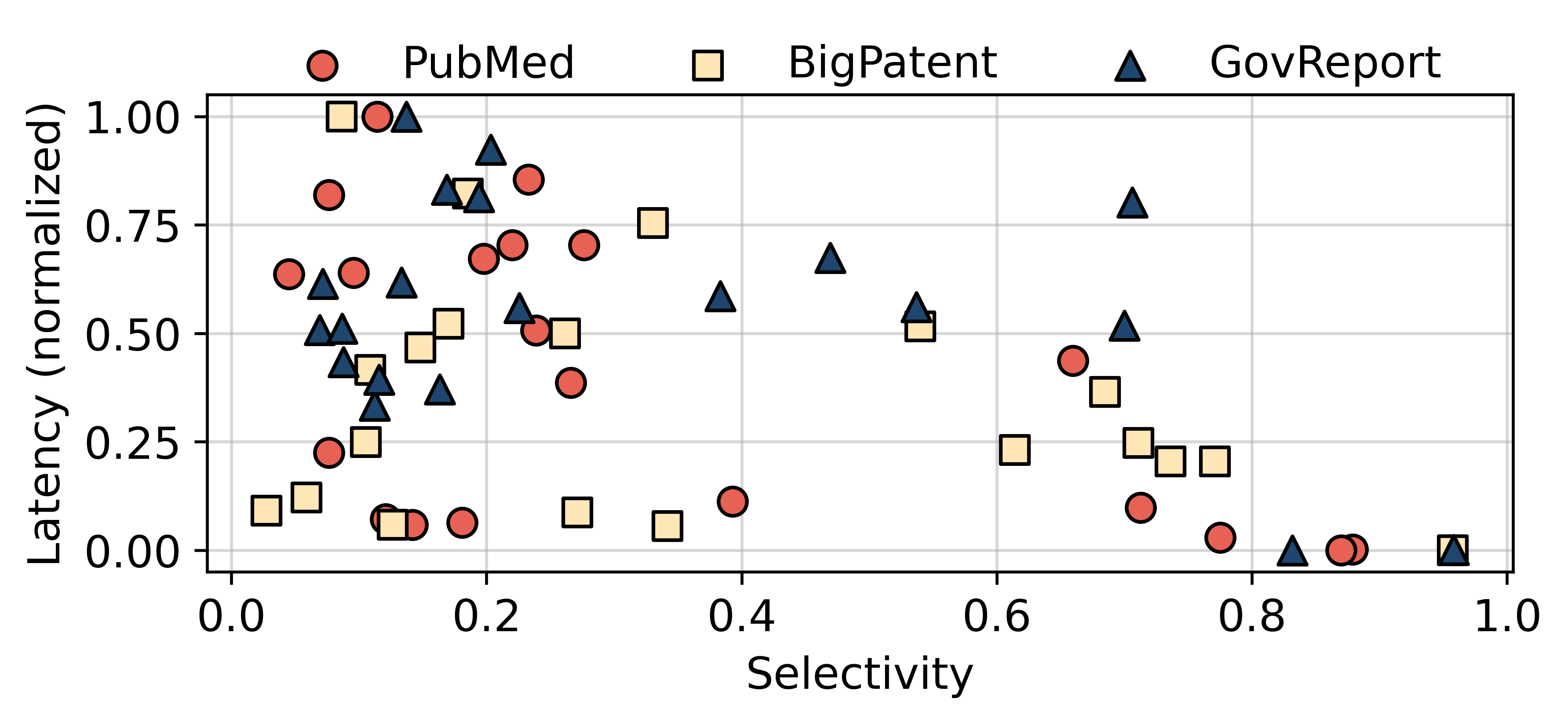}
  \caption{Latency performance (normalized) over different query selectivity.}
  \label{fig:selec}
\end{figure}

A key challenge in data predicates is skewness. While \autoref{fig:acc-selec} shows the relevant accuracy performance,
\autoref{fig:selec} further plots the latency results against selectivity. Queries with higher selectivity achieve greater speedups, as more positive cases form a richer predicate-relevant training set. For low-selectivity queries, performance would be expected to degrade due to the sparse positive data. However, our results reveal a crucial strength: \textsc{ScaleDoc} not only performs well on high-selectivity queries but also delivers evident speedups across low-selectivity scenarios.
This resilience stems from our core contributions. First, the contrastive-based training creates a reliable representation space, capturing discriminative semantics even from limited positive samples. Second, our adaptive cascade mechanisms further enhance the robustness across different data distribution.

\M{
\subsection{Stress-Testing on Complex Queries}

\label{s6.7}

\begin{figure}[t!]
\centering
\includegraphics[width=\linewidth]{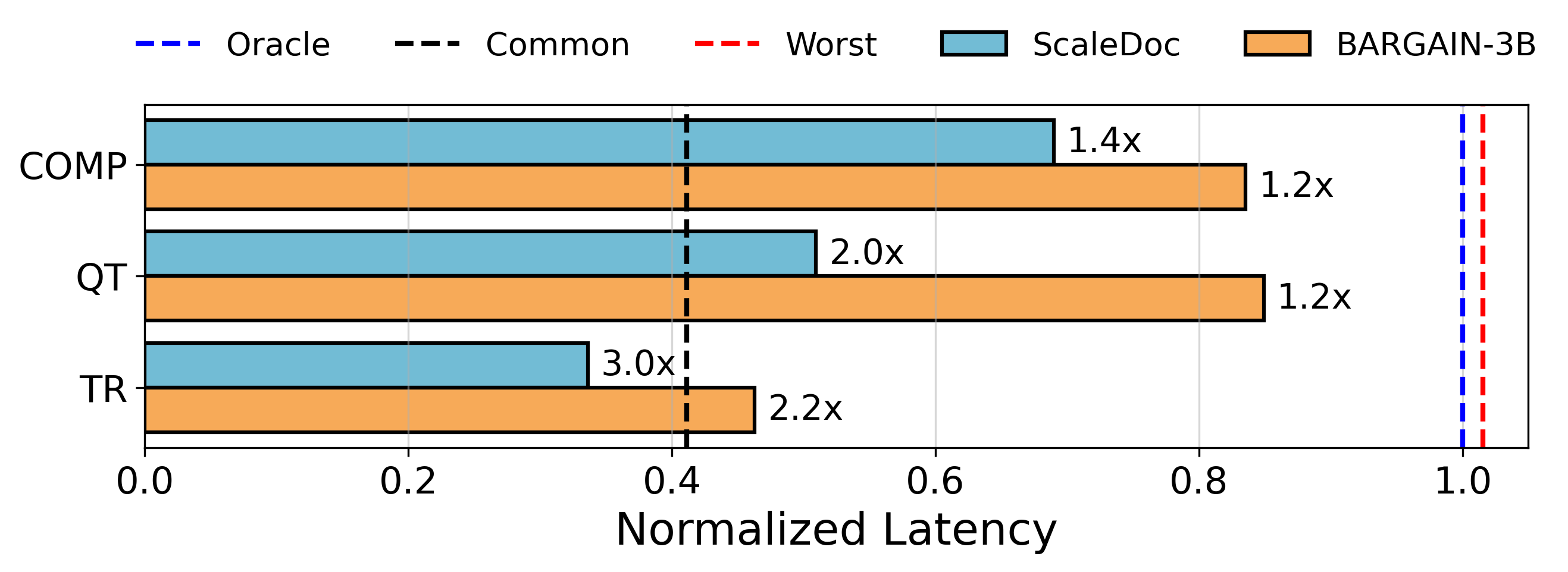}
\caption{\M{Normalized latency on extended queries. \textmd{-- We also label the \textbf{speed-up ratio} compared to the Oracle. \textsc{Common} denotes \textsc{ScaleDoc}'s average latency of existing common queries from PubMed. \textsc{Worst} represents the theoretical worst-case for \textsc{ScaleDoc}   (strictly 100\% target accuracy, no reduction of Oracle).}}}
\label{fig:stress}
\end{figure}

With three real-word document collections and carefully curated queries, our main evaluation covers diverse common predicates such as topic classification, standpoint analysis, fact-checking and attribute verification.
To further assess robustness beyond them, we stress-test \textsc{ScaleDoc} on more complex queries where embeddings may lack sufficient signals. We categorize these complex queries into three types:
\begin{itemize}[leftmargin=1.5em,itemsep=2pt,topsep=2pt]
\item \textbf{Implicit Text Reasoning (\textsc{TR})}: Queries requiring inference on information not explicitly stated in the text.

\emph{Example.} For patent documents, the query could be: ``Does this document describe an invention requiring a highly educated person to operate?''

\item \textbf{Quantitative Analysis (\textsc{QT}):} Tasks with numerical extraction and arithmetic conditions.

\emph{Example:} Medical papers often report p-value tests.  The query could be: ``Does the paper report a p-value less than 0.05?''

\item \textbf{Composite Predicates (\textsc{COMP}):} Queries combining multiple conditions with logical conjunctions.

\emph{Example:} For patent documents, the query could be: ``Is this document from a scientific institution and suitable for common usage in daily life?''
\end{itemize}

We curate 10 new queries across these categories and set the target accuracy as 0.9.
As shown in \autoref{fig:stress}, \textsc{ScaleDoc} consistently outperforms Oracle and BARGAIN-3B, achieving 1.4$\times$--3.0$\times$ speedups.
While effective, our speedups for Quantitative Analysis and Composite Predicates are lower compared to other queries.
This confirms that semantic embeddings are less effective in capturing symbolic patterns (e.g., numerical values) and complex logic.
However, our system could still mitigate this and achieve speedups with adaptive training and effective cascade.
Even in the theoretical worst case where the proxy filters no documents, the overhead of our lightweight proxy is negligible, compared to the expensive LLM inference.
This further ensures that \textsc{ScaleDoc} remains robust in complex scenarios.

}

\begin{figure}[t!]
  \centering
    \includegraphics[width= 0.9\linewidth]{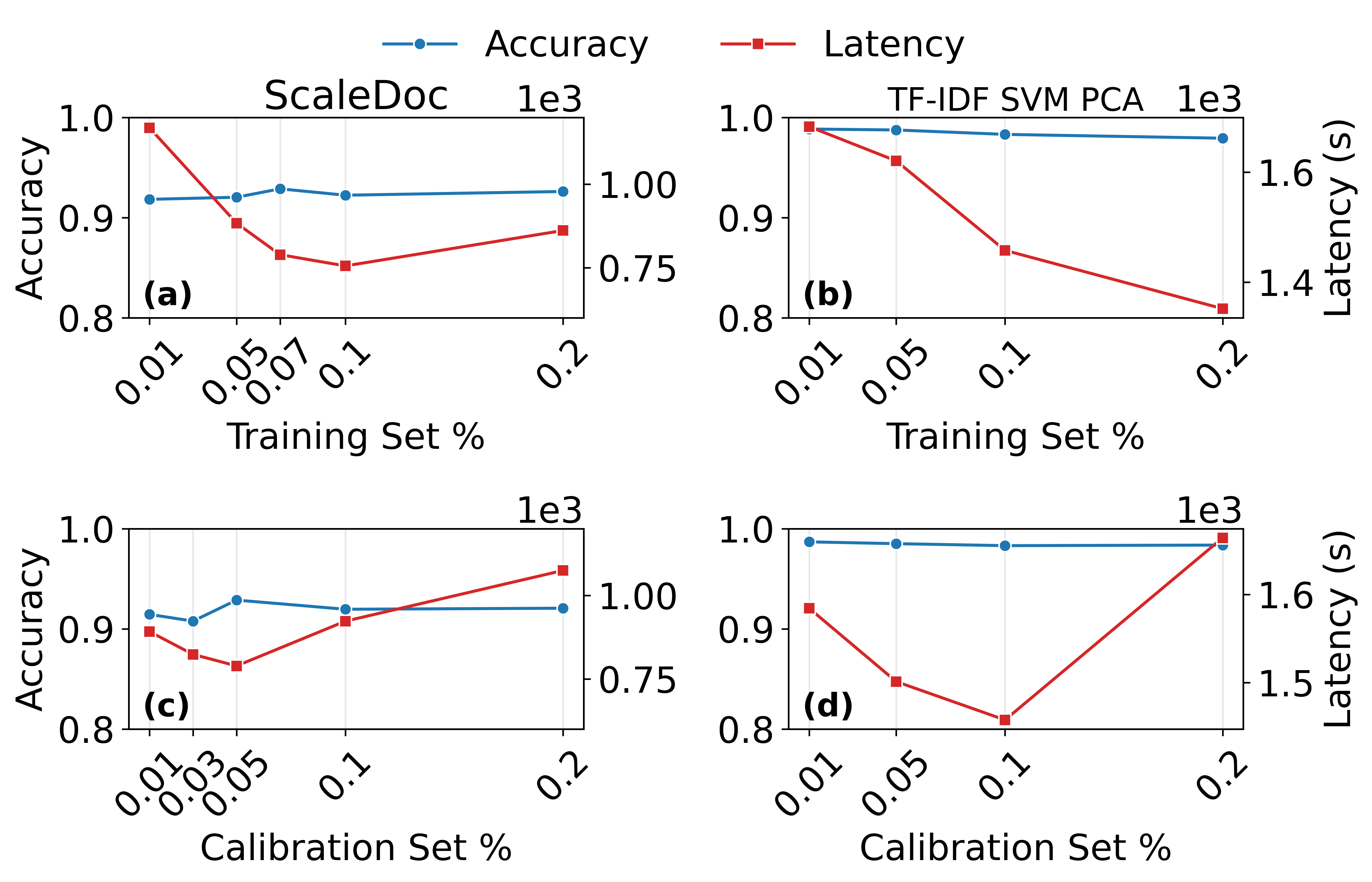}
  \caption{Accuracy and Latency with different hyperparameters -- \textmd{Training Set \% and Calibration Set \% denotes the portion of data sampled from the global.}}
  \label{fig:hyp}
\end{figure}
\subsection{Impact of Hyperparameters}
\label{s7.7}


We demonstrate the impact of different hyperparameters, the sizes of training and calibration sets, on \textsc{ScaleDoc}'s end-to-end performance with a target accuracy of 0.90 (see \autoref{fig:hyp}).
For the training set, larger sizes initially improve the proxy model's ability to learn global features, enhancing efficacy. However, this eventually led to increased labeling overhead and higher latency. The preferred trade-off is empirically found with a training set size between 7\% and 10\%.  Conversely, a traditional PP requires much more labels to train, leading to more LLM oracle invocations. A similar trade-off exists for the calibration set, where sampling 5\% is sufficient to capture global distributions, maximizing end-to-end performance.

\section{Related Work}
\label{s_related}

\vspace{4 pt}
\noindent\textbf{LLMs in Data Systems.} \hspace{0.2cm} Large Language Models (LLMs) has catalyzed a paradigm shift in data analysis, with their remarkable zero-shot capabilities. For instance, numerous studies have explored using LLMs to translate natural language into SQL queries~\cite{li2024using, luoma2025snails, liu-etal-2024-suql}, lowering the barrier to database analytics. LLMs are also integrated as agentic components to orchestrate complex data processing workflows~\cite{shankar2024docetl, 10.1145/3725352, hui2025interact}.  Furthermore, analytical systems such as \textsc{LOTUS}~\cite{10.14778/3749646.3749685} and \textsc{Palimpzest}~\cite{liu2024declarative} formulate \textit{semantic operators}, which provide and integrate several declarative LLM interfaces over unstructured data. Other works target various essential tasks such as multiclass text classification~\cite{kayali2024chorus}, data cleaning~\cite{zhang2024data}, and entity extraction~\cite{hu2025leap}. In parallel, \textsc{ScaleDoc} addresses the foundational and ubiquitous task of semantic predicating. Our goal is to develop a general, scalable, and cost-effective solution for this primitive, which is critical for a wide range of analytical scenarios.

\vspace{4 pt}
\noindent\textbf{Acceleration for Machine Learning Queries.} \hspace{0.2cm} The high computational cost of machine learning (ML) inference, particularly for LLMs, has made acceleration a critical area of research. 
Some works focus on hardware optimizations. For instance, \textsc{PagedAttention}~\cite{kwon2023efficient} and \textsc{FlashAttention}~\cite{dao2022flashattention} improve performance by optimizing the attention mechanism and memory access at the GPU level. Other scheduling approaches, such as \textsc{DistServe}~\cite{10.5555/3691938.3691949} and \textsc{Sarathi}~\cite{agrawal2024taming}, optimize throughput via disaggregated phases or chunked prefilling. Our work is orthogonal to these approaches, focusing on a higher layer of the analytical system stack.

\one{
Additionally, KV caching mechanisms for LLMs are cost-effective by reusing shared prefixes, which is common in multi-turn dialogues or specific context comprehension ~\cite{zheng2024sglang,feng2024adakv}. 
However, our workload requires traversing an enormous and diverse set of
documents. Since these documents are distinct and lack significant
shared prefixes, the overall cache reuse is negligible. Besides, extending the KV cache to cover the entire document collection would incur prohibitive memory overhead, rendering this approach infeasible for our use case.}

At the higher system level, some recent works adopt the proxy-cascades workflow. In this paradigm, a lightweight proxy model rapidly filters out easy entities before sending them to a powerful, expensive model. For instance, Probabilistic Predicates (PPs)~\cite{lu2018accelerating, yang2022optimizing} employed simple classifiers as pre-filtering proxies, and \textsc{NoScope}~\cite{kang2017noscope} built a cascade video databases to accelerate object detection. However, these approaches lack generalizability to new workload patterns, requiring labor-intensive manual adaptation.
More recently, systems such as \textsc{FrugalGPT}~\cite{chen2023frugalgptuselargelanguage} and \textsc{Palimpzest}~\cite{liu2024declarative} use smaller LLMs as proxies for more powerful ones (e.g, GPT-4o). But these solutions still depend on the moderate-scale LLM proxies, remaining expensive.
\one{Furthermore, given the ad hoc and diverse workload, directly fine-tuning small LLMs (e.g., 7B-parameter models) as the oracle is infeasible,  due to the substantial computational overhead of online training~\cite{vm2024finetune-cost}.}
Our work addresses these limitations by exploring a more generalizable, lightweight, and automated solution for semantic queries.


\vspace{4 pt}
\noindent\textbf{Semantic Representation.} \hspace{0.3cm}  Semantic representations encode the meaning of text into dense vectors~\cite{ramesh-kashyap-etal-2024-comprehensive}. Early advances, such as \textsc{Sentence-BERT}~\cite{reimers2019sentence} and \textsc{SimCSE}~\cite{gao-etal-2021-simcse}, established the effectiveness of embeddings for semantic tasks. More recently, the focus has shifted to leveraging LLMs to generate high-quality, context-aware embeddings~\cite{llm2vec, li2023towards, lee2024nv}. These models have demonstrated state-of-the-art performance in various downstream applications~\cite{su2023one, ni2022large}. Our work does not propose a new representation model. Instead, we focus on the systems challenge of efficient workflow orchestration and treat the  representation model as a pluggable component.



\section{Conclusion}


In this paper, we introduce \textsc{ScaleDoc}, a novel system designed to efficiently scale LLM-based semantic predicates over large document collections. It decouples the execution into an offline representation phase and a highly optimized online filtering phase. The system's effectiveness stems from two key innovations: (1) a query-aware proxy model to produce reliable predicating scores, and (2) an ad hoc cascade workflow that dynamically determines the data filtering with specified accuracy targets.
Experiments on diverse datasets confirm our design, showing \textsc{ScaleDoc} reduces LLM calls by up to 85\% and accelerates end-to-end performance by more than 2$\times$. Our work highlights the potential for the use of LLMs in large-scale data analysis systems with effective performance.

\clearpage
\onecolumn
\begin{multicols}{2}
   \bibliographystyle{ACM-Reference-Format}
   \bibliography{ref}


\begin{thebibliography}{52}


\ifx \showCODEN    \undefined \def \showCODEN     #1{\unskip}     \fi
\ifx \showISBNx    \undefined \def \showISBNx     #1{\unskip}     \fi
\ifx \showISBNxiii \undefined \def \showISBNxiii  #1{\unskip}     \fi
\ifx \showISSN     \undefined \def \showISSN      #1{\unskip}     \fi
\ifx \showLCCN     \undefined \def \showLCCN      #1{\unskip}     \fi
\ifx \shownote     \undefined \def \shownote      #1{#1}          \fi
\ifx \showarticletitle \undefined \def \showarticletitle #1{#1}   \fi
\ifx \showURL      \undefined \def \showURL       {\relax}        \fi
\providecommand\bibfield[2]{#2}
\providecommand\bibinfo[2]{#2}
\providecommand\natexlab[1]{#1}
\providecommand\showeprint[2][]{arXiv:#2}

\bibitem[Agrawal et~al\mbox{.}(2024)]%
        {agrawal2024taming}
\bibfield{author}{\bibinfo{person}{Amey Agrawal}, \bibinfo{person}{Nitin Kedia}, \bibinfo{person}{Ashish Panwar}, \bibinfo{person}{Jayashree Mohan}, \bibinfo{person}{Nipun Kwatra}, \bibinfo{person}{Bhargav Gulavani}, \bibinfo{person}{Alexey Tumanov}, {and} \bibinfo{person}{Ramachandran Ramjee}.} \bibinfo{year}{2024}\natexlab{}.
\newblock \showarticletitle{Taming $\{$Throughput-Latency$\}$ tradeoff in $\{$LLM$\}$ inference with $\{$Sarathi-Serve$\}$}. In \bibinfo{booktitle}{\emph{18th USENIX Symposium on Operating Systems Design and Implementation (OSDI 24)}}. \bibinfo{pages}{117--134}.
\newblock


\bibitem[Arora et~al\mbox{.}(2023)]%
        {arora2023llm-dataview}
\bibfield{author}{\bibinfo{person}{Simran Arora}, \bibinfo{person}{Brandon Yang}, \bibinfo{person}{Sabri Eyuboglu}, \bibinfo{person}{Avanika Narayan}, \bibinfo{person}{Andrew Hojel}, \bibinfo{person}{Immanuel Trummer}, {and} \bibinfo{person}{Christopher R{\'e}}.} \bibinfo{year}{2023}\natexlab{}.
\newblock \showarticletitle{Language Models Enable Simple Systems for Generating Structured Views of Heterogeneous Data Lakes}.
\newblock \bibinfo{journal}{\emph{Proceedings of the VLDB Endowment}} \bibinfo{volume}{17}, \bibinfo{number}{2} (\bibinfo{year}{2023}), \bibinfo{pages}{92--105}.
\newblock


\bibitem[BehnamGhader et~al\mbox{.}(2024)]%
        {llm2vec}
\bibfield{author}{\bibinfo{person}{Parishad BehnamGhader}, \bibinfo{person}{Vaibhav Adlakha}, \bibinfo{person}{Marius Mosbach}, \bibinfo{person}{Dzmitry Bahdanau}, \bibinfo{person}{Nicolas Chapados}, {and} \bibinfo{person}{Siva Reddy}.} \bibinfo{year}{2024}\natexlab{}.
\newblock \showarticletitle{{LLM2V}ec: Large Language Models Are Secretly Powerful Text Encoders}. In \bibinfo{booktitle}{\emph{First Conference on Language Modeling}}.
\newblock
\urldef\tempurl%
\url{https://openreview.net/forum?id=IW1PR7vEBf}
\showURL{%
\tempurl}


\bibitem[Chen et~al\mbox{.}(2023)]%
        {chen2023frugalgptuselargelanguage}
\bibfield{author}{\bibinfo{person}{Lingjiao Chen}, \bibinfo{person}{Matei Zaharia}, {and} \bibinfo{person}{James Zou}.} \bibinfo{year}{2023}\natexlab{}.
\newblock \bibinfo{title}{FrugalGPT: How to Use Large Language Models While Reducing Cost and Improving Performance}.
\newblock
\showeprint[arxiv]{2305.05176}~[cs.LG]
\urldef\tempurl%
\url{https://arxiv.org/abs/2305.05176}
\showURL{%
\tempurl}


\bibitem[Chen et~al\mbox{.}(2020)]%
        {chen2020simple}
\bibfield{author}{\bibinfo{person}{Ting Chen}, \bibinfo{person}{Simon Kornblith}, \bibinfo{person}{Mohammad Norouzi}, {and} \bibinfo{person}{Geoffrey Hinton}.} \bibinfo{year}{2020}\natexlab{}.
\newblock \showarticletitle{A simple framework for contrastive learning of visual representations}. In \bibinfo{booktitle}{\emph{International conference on machine learning}}. PmLR, \bibinfo{pages}{1597--1607}.
\newblock


\bibitem[Dao et~al\mbox{.}(2022)]%
        {dao2022flashattention}
\bibfield{author}{\bibinfo{person}{Tri Dao}, \bibinfo{person}{Dan Fu}, \bibinfo{person}{Stefano Ermon}, \bibinfo{person}{Atri Rudra}, {and} \bibinfo{person}{Christopher R{\'e}}.} \bibinfo{year}{2022}\natexlab{}.
\newblock \showarticletitle{Flashattention: Fast and memory-efficient exact attention with io-awareness}.
\newblock \bibinfo{journal}{\emph{Advances in neural information processing systems}}  \bibinfo{volume}{35} (\bibinfo{year}{2022}), \bibinfo{pages}{16344--16359}.
\newblock


\bibitem[Dernoncourt and Lee(2017)]%
        {dernoncourt-lee-2017-pubmed}
\bibfield{author}{\bibinfo{person}{Franck Dernoncourt} {and} \bibinfo{person}{Ji~Young Lee}.} \bibinfo{year}{2017}\natexlab{}.
\newblock \showarticletitle{{P}ub{M}ed 200k {RCT}: a Dataset for Sequential Sentence Classification in Medical Abstracts}. In \bibinfo{booktitle}{\emph{Proceedings of the Eighth International Joint Conference on Natural Language Processing (Volume 2: Short Papers)}}, \bibfield{editor}{\bibinfo{person}{Greg Kondrak} {and} \bibinfo{person}{Taro Watanabe}} (Eds.). \bibinfo{publisher}{Asian Federation of Natural Language Processing}, \bibinfo{address}{Taipei, Taiwan}, \bibinfo{pages}{308--313}.
\newblock
\urldef\tempurl%
\url{https://aclanthology.org/I17-2052/}
\showURL{%
\tempurl}


\bibitem[Feng et~al\mbox{.}(2024)]%
        {feng2024adakv}
\bibfield{author}{\bibinfo{person}{Yuan Feng}, \bibinfo{person}{Junlin Lv}, \bibinfo{person}{Yukun Cao}, \bibinfo{person}{Xike Xie}, {and} \bibinfo{person}{S~Kevin Zhou}.} \bibinfo{year}{2024}\natexlab{}.
\newblock \showarticletitle{Ada-kv: Optimizing kv cache eviction by adaptive budget allocation for efficient llm inference}.
\newblock \bibinfo{journal}{\emph{arXiv preprint arXiv:2407.11550}} (\bibinfo{year}{2024}).
\newblock


\bibitem[Gao et~al\mbox{.}(2021)]%
        {gao-etal-2021-simcse}
\bibfield{author}{\bibinfo{person}{Tianyu Gao}, \bibinfo{person}{Xingcheng Yao}, {and} \bibinfo{person}{Danqi Chen}.} \bibinfo{year}{2021}\natexlab{}.
\newblock \showarticletitle{{S}im{CSE}: Simple Contrastive Learning of Sentence Embeddings}. In \bibinfo{booktitle}{\emph{Proceedings of the 2021 Conference on Empirical Methods in Natural Language Processing}}, \bibfield{editor}{\bibinfo{person}{Marie-Francine Moens}, \bibinfo{person}{Xuanjing Huang}, \bibinfo{person}{Lucia Specia}, {and} \bibinfo{person}{Scott Wen-tau Yih}} (Eds.). \bibinfo{publisher}{Association for Computational Linguistics}, \bibinfo{address}{Online and Punta Cana, Dominican Republic}, \bibinfo{pages}{6894--6910}.
\newblock
\href{https://doi.org/10.18653/v1/2021.emnlp-main.552}{doi:\nolinkurl{10.18653/v1/2021.emnlp-main.552}}


\bibitem[He et~al\mbox{.}(2020)]%
        {he2020momentum}
\bibfield{author}{\bibinfo{person}{Kaiming He}, \bibinfo{person}{Haoqi Fan}, \bibinfo{person}{Yuxin Wu}, \bibinfo{person}{Saining Xie}, {and} \bibinfo{person}{Ross Girshick}.} \bibinfo{year}{2020}\natexlab{}.
\newblock \showarticletitle{Momentum contrast for unsupervised visual representation learning}. In \bibinfo{booktitle}{\emph{Proceedings of the IEEE/CVF conference on computer vision and pattern recognition}}. \bibinfo{pages}{9729--9738}.
\newblock


\bibitem[Hu et~al\mbox{.}(2025)]%
        {hu2025leap}
\bibfield{author}{\bibinfo{person}{Chuxuan Hu}, \bibinfo{person}{Austin Peters}, {and} \bibinfo{person}{Daniel Kang}.} \bibinfo{year}{2025}\natexlab{}.
\newblock \showarticletitle{LEAP: LLM-powered End-to-end Automatic Library for Processing Social Science Queries on Unstructured Data}.
\newblock \bibinfo{journal}{\emph{arXiv preprint arXiv:2501.03892}} (\bibinfo{year}{2025}).
\newblock


\bibitem[Huang et~al\mbox{.}(2020)]%
        {huang2020embedding-ret}
\bibfield{author}{\bibinfo{person}{Jui-Ting Huang}, \bibinfo{person}{Ashish Sharma}, \bibinfo{person}{Shuying Sun}, \bibinfo{person}{Li Xia}, \bibinfo{person}{David Zhang}, \bibinfo{person}{Philip Pronin}, \bibinfo{person}{Janani Padmanabhan}, \bibinfo{person}{Giuseppe Ottaviano}, {and} \bibinfo{person}{Linjun Yang}.} \bibinfo{year}{2020}\natexlab{}.
\newblock \showarticletitle{Embedding-based retrieval in facebook search}. In \bibinfo{booktitle}{\emph{Proceedings of the 26th ACM SIGKDD International Conference on Knowledge Discovery \& Data Mining}}. \bibinfo{pages}{2553--2561}.
\newblock


\bibitem[Huang et~al\mbox{.}(2021)]%
        {huang-etal-2021-efficient}
\bibfield{author}{\bibinfo{person}{Luyang Huang}, \bibinfo{person}{Shuyang Cao}, \bibinfo{person}{Nikolaus Parulian}, \bibinfo{person}{Heng Ji}, {and} \bibinfo{person}{Lu Wang}.} \bibinfo{year}{2021}\natexlab{}.
\newblock \showarticletitle{Efficient Attentions for Long Document Summarization}. In \bibinfo{booktitle}{\emph{Proceedings of the 2021 Conference of the North American Chapter of the Association for Computational Linguistics: Human Language Technologies}}. \bibinfo{publisher}{Association for Computational Linguistics}, \bibinfo{address}{Online}, \bibinfo{pages}{1419--1436}.
\newblock
\href{https://doi.org/10.18653/v1/2021.naacl-main.112}{doi:\nolinkurl{10.18653/v1/2021.naacl-main.112}}


\bibitem[Hui et~al\mbox{.}(2025a)]%
        {hui2025interact}
\bibfield{author}{\bibinfo{person}{Yulong Hui}, \bibinfo{person}{Chao Chen}, \bibinfo{person}{Zhihang Fu}, \bibinfo{person}{Yihao Liu}, \bibinfo{person}{Jieping Ye}, {and} \bibinfo{person}{Huanchen Zhang}.} \bibinfo{year}{2025}\natexlab{a}.
\newblock \showarticletitle{Interact-RAG: Reason and Interact with the Corpus, Beyond Black-Box Retrieval}.
\newblock \bibinfo{journal}{\emph{arXiv preprint arXiv:2510.27566}} (\bibinfo{year}{2025}).
\newblock


\bibitem[Hui et~al\mbox{.}(2025b)]%
        {hui2025okralong}
\bibfield{author}{\bibinfo{person}{Yulong Hui}, \bibinfo{person}{Yihao Liu}, \bibinfo{person}{Yao Lu}, {and} \bibinfo{person}{Huanchen Zhang}.} \bibinfo{year}{2025}\natexlab{b}.
\newblock \showarticletitle{Okralong: A flexible retrieval-augmented framework for long-text query processing}.
\newblock \bibinfo{journal}{\emph{arXiv preprint arXiv:2503.02603}} (\bibinfo{year}{2025}).
\newblock


\bibitem[Hui et~al\mbox{.}({[n.\,d.]})]%
        {huiuda}
\bibfield{author}{\bibinfo{person}{Yulong Hui}, \bibinfo{person}{Yao Lu}, {and} \bibinfo{person}{Huanchen Zhang}.} \bibinfo{year}{[n.\,d.]}\natexlab{}.
\newblock \showarticletitle{UDA: A Benchmark Suite for Retrieval Augmented Generation in Real-World Document Analysis}. In \bibinfo{booktitle}{\emph{The Thirty-eight Conference on Neural Information Processing Systems Datasets and Benchmarks Track}}.
\newblock


\bibitem[Jiang et~al\mbox{.}(2023)]%
        {jiang2023mistral7b}
\bibfield{author}{\bibinfo{person}{Albert~Q. Jiang}, \bibinfo{person}{Alexandre Sablayrolles}, \bibinfo{person}{Arthur Mensch}, \bibinfo{person}{Chris Bamford}, \bibinfo{person}{Devendra~Singh Chaplot}, \bibinfo{person}{Diego de~las Casas}, \bibinfo{person}{Florian Bressand}, \bibinfo{person}{Gianna Lengyel}, \bibinfo{person}{Guillaume Lample}, \bibinfo{person}{Lucile Saulnier}, \bibinfo{person}{Lélio~Renard Lavaud}, \bibinfo{person}{Marie-Anne Lachaux}, \bibinfo{person}{Pierre Stock}, \bibinfo{person}{Teven~Le Scao}, \bibinfo{person}{Thibaut Lavril}, \bibinfo{person}{Thomas Wang}, \bibinfo{person}{Timothée Lacroix}, {and} \bibinfo{person}{William~El Sayed}.} \bibinfo{year}{2023}\natexlab{}.
\newblock \bibinfo{title}{Mistral 7B}.
\newblock
\showeprint[arxiv]{2310.06825}~[cs.CL]
\urldef\tempurl%
\url{https://arxiv.org/abs/2310.06825}
\showURL{%
\tempurl}


\bibitem[Kang et~al\mbox{.}(2017)]%
        {kang2017noscope}
\bibfield{author}{\bibinfo{person}{Daniel Kang}, \bibinfo{person}{John Emmons}, \bibinfo{person}{Firas Abuzaid}, \bibinfo{person}{Peter Bailis}, {and} \bibinfo{person}{Matei Zaharia}.} \bibinfo{year}{2017}\natexlab{}.
\newblock \showarticletitle{NoScope: Optimizing Neural Network Queries over Video at Scale}.
\newblock \bibinfo{journal}{\emph{Proceedings of the VLDB Endowment}} \bibinfo{volume}{10}, \bibinfo{number}{11} (\bibinfo{year}{2017}).
\newblock


\bibitem[Kang et~al\mbox{.}(2020)]%
        {10.14778/3407790.3407804}
\bibfield{author}{\bibinfo{person}{Daniel Kang}, \bibinfo{person}{Edward Gan}, \bibinfo{person}{Peter Bailis}, \bibinfo{person}{Tatsunori Hashimoto}, {and} \bibinfo{person}{Matei Zaharia}.} \bibinfo{year}{2020}\natexlab{}.
\newblock \showarticletitle{Approximate selection with guarantees using proxies}.
\newblock \bibinfo{journal}{\emph{Proc. VLDB Endow.}} \bibinfo{volume}{13}, \bibinfo{number}{12} (\bibinfo{date}{July} \bibinfo{year}{2020}), \bibinfo{pages}{1990–2003}.
\newblock
\showISSN{2150-8097}
\href{https://doi.org/10.14778/3407790.3407804}{doi:\nolinkurl{10.14778/3407790.3407804}}


\bibitem[Karpukhin et~al\mbox{.}({[n.\,d.]})]%
        {karpukhin2020dense}
\bibfield{author}{\bibinfo{person}{Vladimir Karpukhin}, \bibinfo{person}{Barlas Oguz}, \bibinfo{person}{Sewon Min}, \bibinfo{person}{Ledell Wu}, \bibinfo{person}{Sergey Edunov}, \bibinfo{person}{Danqi Chen}, {and} \bibinfo{person}{Wen-tau Yih}.} \bibinfo{year}{[n.\,d.]}\natexlab{}.
\newblock \showarticletitle{Dense Passage Retrieval for Open-Domain Question Answering.}
\newblock


\bibitem[Kayali et~al\mbox{.}(2024)]%
        {kayali2024chorus}
\bibfield{author}{\bibinfo{person}{Moe Kayali}, \bibinfo{person}{Anton Lykov}, \bibinfo{person}{Ilias Fountalis}, \bibinfo{person}{Nikolaos Vasiloglou}, \bibinfo{person}{Dan Olteanu}, {and} \bibinfo{person}{Dan Suciu}.} \bibinfo{year}{2024}\natexlab{}.
\newblock \showarticletitle{Chorus: Foundation Models for Unified Data Discovery and Exploration}.
\newblock \bibinfo{journal}{\emph{Proceedings of the VLDB Endowment}} \bibinfo{volume}{17}, \bibinfo{number}{8} (\bibinfo{year}{2024}), \bibinfo{pages}{2104--2114}.
\newblock


\bibitem[Khosla et~al\mbox{.}(2020)]%
        {khosla2020supervised}
\bibfield{author}{\bibinfo{person}{Prannay Khosla}, \bibinfo{person}{Piotr Teterwak}, \bibinfo{person}{Chen Wang}, \bibinfo{person}{Aaron Sarna}, \bibinfo{person}{Yonglong Tian}, \bibinfo{person}{Phillip Isola}, \bibinfo{person}{Aaron Maschinot}, \bibinfo{person}{Ce Liu}, {and} \bibinfo{person}{Dilip Krishnan}.} \bibinfo{year}{2020}\natexlab{}.
\newblock \showarticletitle{Supervised contrastive learning}.
\newblock \bibinfo{journal}{\emph{Advances in neural information processing systems}}  \bibinfo{volume}{33} (\bibinfo{year}{2020}), \bibinfo{pages}{18661--18673}.
\newblock


\bibitem[Kwon et~al\mbox{.}(2023)]%
        {kwon2023efficient}
\bibfield{author}{\bibinfo{person}{Woosuk Kwon}, \bibinfo{person}{Zhuohan Li}, \bibinfo{person}{Siyuan Zhuang}, \bibinfo{person}{Ying Sheng}, \bibinfo{person}{Lianmin Zheng}, \bibinfo{person}{Cody~Hao Yu}, \bibinfo{person}{Joseph Gonzalez}, \bibinfo{person}{Hao Zhang}, {and} \bibinfo{person}{Ion Stoica}.} \bibinfo{year}{2023}\natexlab{}.
\newblock \showarticletitle{Efficient memory management for large language model serving with pagedattention}. In \bibinfo{booktitle}{\emph{Proceedings of the 29th Symposium on Operating Systems Principles}}. \bibinfo{pages}{611--626}.
\newblock


\bibitem[Lee et~al\mbox{.}(2024)]%
        {lee2024nv}
\bibfield{author}{\bibinfo{person}{Chankyu Lee}, \bibinfo{person}{Rajarshi Roy}, \bibinfo{person}{Mengyao Xu}, \bibinfo{person}{Jonathan Raiman}, \bibinfo{person}{Mohammad Shoeybi}, \bibinfo{person}{Bryan Catanzaro}, {and} \bibinfo{person}{Wei Ping}.} \bibinfo{year}{2024}\natexlab{}.
\newblock \showarticletitle{Nv-embed: Improved techniques for training llms as generalist embedding models}.
\newblock \bibinfo{journal}{\emph{arXiv preprint arXiv:2405.17428}} (\bibinfo{year}{2024}).
\newblock


\bibitem[Li and Xie(2024)]%
        {li2024using}
\bibfield{author}{\bibinfo{person}{Zhenwen Li} {and} \bibinfo{person}{Tao Xie}.} \bibinfo{year}{2024}\natexlab{}.
\newblock \showarticletitle{Using LLM to select the right SQL Query from candidates}.
\newblock \bibinfo{journal}{\emph{arXiv preprint arXiv:2401.02115}} (\bibinfo{year}{2024}).
\newblock


\bibitem[Li et~al\mbox{.}(2023)]%
        {li2023towards}
\bibfield{author}{\bibinfo{person}{Zehan Li}, \bibinfo{person}{Xin Zhang}, \bibinfo{person}{Yanzhao Zhang}, \bibinfo{person}{Dingkun Long}, \bibinfo{person}{Pengjun Xie}, {and} \bibinfo{person}{Meishan Zhang}.} \bibinfo{year}{2023}\natexlab{}.
\newblock \showarticletitle{Towards general text embeddings with multi-stage contrastive learning}.
\newblock \bibinfo{journal}{\emph{arXiv preprint arXiv:2308.03281}} (\bibinfo{year}{2023}).
\newblock


\bibitem[Liu et~al\mbox{.}(2024a)]%
        {liu2024declarative}
\bibfield{author}{\bibinfo{person}{Chunwei Liu}, \bibinfo{person}{Matthew Russo}, \bibinfo{person}{Michael Cafarella}, \bibinfo{person}{Lei Cao}, \bibinfo{person}{Peter~Baille Chen}, \bibinfo{person}{Zui Chen}, \bibinfo{person}{Michael Franklin}, \bibinfo{person}{Tim Kraska}, \bibinfo{person}{Samuel Madden}, {and} \bibinfo{person}{Gerardo Vitagliano}.} \bibinfo{year}{2024}\natexlab{a}.
\newblock \showarticletitle{A declarative system for optimizing ai workloads}.
\newblock \bibinfo{journal}{\emph{arXiv preprint arXiv:2405.14696}} (\bibinfo{year}{2024}).
\newblock


\bibitem[Liu et~al\mbox{.}(2024b)]%
        {liu-etal-2024-suql}
\bibfield{author}{\bibinfo{person}{Shicheng Liu}, \bibinfo{person}{Jialiang Xu}, \bibinfo{person}{Wesley Tjangnaka}, \bibinfo{person}{Sina Semnani}, \bibinfo{person}{Chen Yu}, {and} \bibinfo{person}{Monica Lam}.} \bibinfo{year}{2024}\natexlab{b}.
\newblock \showarticletitle{{SUQL}: Conversational Search over Structured and Unstructured Data with Large Language Models}. In \bibinfo{booktitle}{\emph{Findings of the Association for Computational Linguistics: NAACL 2024}}, \bibfield{editor}{\bibinfo{person}{Kevin Duh}, \bibinfo{person}{Helena Gomez}, {and} \bibinfo{person}{Steven Bethard}} (Eds.). \bibinfo{publisher}{Association for Computational Linguistics}, \bibinfo{address}{Mexico City, Mexico}, \bibinfo{pages}{4535--4555}.
\newblock
\href{https://doi.org/10.18653/v1/2024.findings-naacl.283}{doi:\nolinkurl{10.18653/v1/2024.findings-naacl.283}}


\bibitem[Lu et~al\mbox{.}(2018)]%
        {lu2018accelerating}
\bibfield{author}{\bibinfo{person}{Yao Lu}, \bibinfo{person}{Aakanksha Chowdhery}, \bibinfo{person}{Srikanth Kandula}, {and} \bibinfo{person}{Surajit Chaudhuri}.} \bibinfo{year}{2018}\natexlab{}.
\newblock \showarticletitle{Accelerating machine learning inference with probabilistic predicates}. In \bibinfo{booktitle}{\emph{Proceedings of the 2018 International Conference on Management of Data}}. \bibinfo{pages}{1493--1508}.
\newblock


\bibitem[Luoma and Kumar(2025)]%
        {luoma2025snails}
\bibfield{author}{\bibinfo{person}{Kyle Luoma} {and} \bibinfo{person}{Arun Kumar}.} \bibinfo{year}{2025}\natexlab{}.
\newblock \showarticletitle{SNAILS: Schema Naming Assessments for Improved LLM-Based SQL Inference}.
\newblock \bibinfo{journal}{\emph{Proceedings of the ACM on Management of Data}} \bibinfo{volume}{3}, \bibinfo{number}{1} (\bibinfo{year}{2025}), \bibinfo{pages}{1--26}.
\newblock


\bibitem[Ni et~al\mbox{.}(2022)]%
        {ni2022large}
\bibfield{author}{\bibinfo{person}{Jianmo Ni}, \bibinfo{person}{Chen Qu}, \bibinfo{person}{Jing Lu}, \bibinfo{person}{Zhuyun Dai}, \bibinfo{person}{Gustavo~Hernandez Abrego}, \bibinfo{person}{Ji Ma}, \bibinfo{person}{Vincent Zhao}, \bibinfo{person}{Yi Luan}, \bibinfo{person}{Keith Hall}, \bibinfo{person}{Ming-Wei Chang}, {et~al\mbox{.}}} \bibinfo{year}{2022}\natexlab{}.
\newblock \showarticletitle{Large Dual Encoders Are Generalizable Retrievers}. In \bibinfo{booktitle}{\emph{Proceedings of the 2022 Conference on Empirical Methods in Natural Language Processing}}. \bibinfo{pages}{9844--9855}.
\newblock


\bibitem[OpenAI et~al\mbox{.}(2024)]%
        {openai2024gpt4ocard}
\bibfield{author}{\bibinfo{person}{OpenAI}, \bibinfo{person}{Aaron Hurst}, \bibinfo{person}{Adam Lerer}, \bibinfo{person}{Adam~P. Goucher}, \bibinfo{person}{Adam Perelman}, \bibinfo{person}{Aditya Ramesh}, \bibinfo{person}{Aidan Clark}, \bibinfo{person}{AJ Ostrow}, \bibinfo{person}{Akila Welihinda}, \bibinfo{person}{Alan Hayes}, \bibinfo{person}{Alec Radford}, \bibinfo{person}{Aleksander Mądry}, \bibinfo{person}{Alex Baker-Whitcomb}, \bibinfo{person}{Alex Beutel}, \bibinfo{person}{Alex Borzunov}, \bibinfo{person}{Alex Carney}, \bibinfo{person}{Alex Chow}, \bibinfo{person}{Alex Kirillov}, \bibinfo{person}{Alex Nichol}, \bibinfo{person}{Alex Paino}, \bibinfo{person}{Alex Renzin}, \bibinfo{person}{Alex~Tachard Passos}, \bibinfo{person}{Alexander Kirillov}, \bibinfo{person}{Alexi Christakis}, \bibinfo{person}{Alexis Conneau}, \bibinfo{person}{Ali Kamali}, \bibinfo{person}{Allan Jabri}, \bibinfo{person}{Allison Moyer}, \bibinfo{person}{Allison Tam}, \bibinfo{person}{Amadou Crookes}, \bibinfo{person}{Amin
  Tootoochian}, \bibinfo{person}{Amin Tootoonchian}, \bibinfo{person}{Ananya Kumar}, \bibinfo{person}{Andrea Vallone}, \bibinfo{person}{Andrej Karpathy}, \bibinfo{person}{Andrew Braunstein}, \bibinfo{person}{Andrew Cann}, {and} \bibinfo{person}{Andrew Codispoti}.} \bibinfo{year}{2024}\natexlab{}.
\newblock \bibinfo{title}{GPT-4o System Card}.
\newblock
\showeprint[arxiv]{2410.21276}~[cs.CL]
\urldef\tempurl%
\url{https://arxiv.org/abs/2410.21276}
\showURL{%
\tempurl}


\bibitem[Patel et~al\mbox{.}(2024)]%
        {patel2024lotus}
\bibfield{author}{\bibinfo{person}{Liana Patel}, \bibinfo{person}{Siddharth Jha}, \bibinfo{person}{Carlos Guestrin}, {and} \bibinfo{person}{Matei Zaharia}.} \bibinfo{year}{2024}\natexlab{}.
\newblock \showarticletitle{Lotus: Enabling semantic queries with llms over tables of unstructured and structured data}.
\newblock \bibinfo{journal}{\emph{arXiv preprint arXiv:2407.11418}} (\bibinfo{year}{2024}).
\newblock


\bibitem[Patel et~al\mbox{.}(2025)]%
        {10.14778/3749646.3749685}
\bibfield{author}{\bibinfo{person}{Liana Patel}, \bibinfo{person}{Siddharth Jha}, \bibinfo{person}{Melissa Pan}, \bibinfo{person}{Harshit Gupta}, \bibinfo{person}{Parth Asawa}, \bibinfo{person}{Carlos Guestrin}, {and} \bibinfo{person}{Matei Zaharia}.} \bibinfo{year}{2025}\natexlab{}.
\newblock \showarticletitle{Semantic Operators and Their Optimization: Enabling LLM-Based Data Processing with Accuracy Guarantees in LOTUS}.
\newblock \bibinfo{journal}{\emph{Proc. VLDB Endow.}} \bibinfo{volume}{18}, \bibinfo{number}{11} (\bibinfo{date}{July} \bibinfo{year}{2025}), \bibinfo{pages}{4171–4184}.
\newblock
\showISSN{2150-8097}
\href{https://doi.org/10.14778/3749646.3749685}{doi:\nolinkurl{10.14778/3749646.3749685}}


\bibitem[Pedregosa et~al\mbox{.}(2011)]%
        {pedregosa2011scikit}
\bibfield{author}{\bibinfo{person}{Fabian Pedregosa}, \bibinfo{person}{Ga{\"e}l Varoquaux}, \bibinfo{person}{Alexandre Gramfort}, \bibinfo{person}{Vincent Michel}, \bibinfo{person}{Bertrand Thirion}, \bibinfo{person}{Olivier Grisel}, \bibinfo{person}{Mathieu Blondel}, \bibinfo{person}{Peter Prettenhofer}, \bibinfo{person}{Ron Weiss}, \bibinfo{person}{Vincent Dubourg}, {et~al\mbox{.}}} \bibinfo{year}{2011}\natexlab{}.
\newblock \showarticletitle{scikit-learn: Machine learning in Python}.
\newblock \bibinfo{journal}{\emph{the Journal of machine Learning research}}  \bibinfo{volume}{12} (\bibinfo{year}{2011}), \bibinfo{pages}{2825--2830}.
\newblock


\bibitem[Ramesh~Kashyap et~al\mbox{.}(2024)]%
        {ramesh-kashyap-etal-2024-comprehensive}
\bibfield{author}{\bibinfo{person}{Abhinav Ramesh~Kashyap}, \bibinfo{person}{Thanh-Tung Nguyen}, \bibinfo{person}{Viktor Schlegel}, \bibinfo{person}{Stefan Winkler}, \bibinfo{person}{See-Kiong Ng}, {and} \bibinfo{person}{Soujanya Poria}.} \bibinfo{year}{2024}\natexlab{}.
\newblock \showarticletitle{A Comprehensive Survey of Sentence Representations: From the {BERT} Epoch to the {CHATGPT} Era and Beyond}. In \bibinfo{booktitle}{\emph{Proceedings of the 18th Conference of the European Chapter of the Association for Computational Linguistics (Volume 1: Long Papers)}}, \bibfield{editor}{\bibinfo{person}{Yvette Graham} {and} \bibinfo{person}{Matthew Purver}} (Eds.). \bibinfo{publisher}{Association for Computational Linguistics}, \bibinfo{address}{St. Julian{'}s, Malta}, \bibinfo{pages}{1738--1751}.
\newblock
\urldef\tempurl%
\url{https://aclanthology.org/2024.eacl-long.104/}
\showURL{%
\tempurl}


\bibitem[Reimers and Gurevych(2019)]%
        {reimers2019sentence}
\bibfield{author}{\bibinfo{person}{Nils Reimers} {and} \bibinfo{person}{Iryna Gurevych}.} \bibinfo{year}{2019}\natexlab{}.
\newblock \showarticletitle{Sentence-BERT: Sentence Embeddings using Siamese BERT-Networks}. In \bibinfo{booktitle}{\emph{Proceedings of the 2019 Conference on Empirical Methods in Natural Language Processing and the 9th International Joint Conference on Natural Language Processing (EMNLP-IJCNLP)}}. \bibinfo{pages}{3982--3992}.
\newblock


\bibitem[Salazar-D{\'\i}az et~al\mbox{.}(2024)]%
        {salazar2024inferdb}
\bibfield{author}{\bibinfo{person}{Ricardo Salazar-D{\'\i}az}, \bibinfo{person}{Boris Glavic}, {and} \bibinfo{person}{Tilmann Rabl}.} \bibinfo{year}{2024}\natexlab{}.
\newblock \showarticletitle{Inferdb: In-database machine learning inference using indexes}.
\newblock \bibinfo{journal}{\emph{Proceedings of the VLDB Endowment}} \bibinfo{volume}{17}, \bibinfo{number}{8} (\bibinfo{year}{2024}), \bibinfo{pages}{1830--1842}.
\newblock


\bibitem[Sanh et~al\mbox{.}(2019)]%
        {sanh2019distilbert}
\bibfield{author}{\bibinfo{person}{Victor Sanh}, \bibinfo{person}{Lysandre Debut}, \bibinfo{person}{Julien Chaumond}, {and} \bibinfo{person}{Thomas Wolf}.} \bibinfo{year}{2019}\natexlab{}.
\newblock \showarticletitle{DistilBERT, a distilled version of BERT: smaller, faster, cheaper and lighter}.
\newblock \bibinfo{journal}{\emph{arXiv preprint arXiv:1910.01108}} (\bibinfo{year}{2019}).
\newblock


\bibitem[Shankar et~al\mbox{.}(2024)]%
        {shankar2024docetl}
\bibfield{author}{\bibinfo{person}{Shreya Shankar}, \bibinfo{person}{Tristan Chambers}, \bibinfo{person}{Tarak Shah}, \bibinfo{person}{Aditya~G Parameswaran}, {and} \bibinfo{person}{Eugene Wu}.} \bibinfo{year}{2024}\natexlab{}.
\newblock \showarticletitle{DocETL: Agentic Query Rewriting and Evaluation for Complex Document Processing}.
\newblock \bibinfo{journal}{\emph{arXiv preprint arXiv:2410.12189}} (\bibinfo{year}{2024}).
\newblock


\bibitem[Sharma et~al\mbox{.}(2019)]%
        {sharma-etal-2019-bigpatent}
\bibfield{author}{\bibinfo{person}{Eva Sharma}, \bibinfo{person}{Chen Li}, {and} \bibinfo{person}{Lu Wang}.} \bibinfo{year}{2019}\natexlab{}.
\newblock \showarticletitle{{BIGPATENT}: A Large-Scale Dataset for Abstractive and Coherent Summarization}. In \bibinfo{booktitle}{\emph{Proceedings of the 57th Annual Meeting of the Association for Computational Linguistics}}, \bibfield{editor}{\bibinfo{person}{Anna Korhonen}, \bibinfo{person}{David Traum}, {and} \bibinfo{person}{Llu{\'i}s M{\`a}rquez}} (Eds.). \bibinfo{publisher}{Association for Computational Linguistics}, \bibinfo{address}{Florence, Italy}, \bibinfo{pages}{2204--2213}.
\newblock
\href{https://doi.org/10.18653/v1/P19-1212}{doi:\nolinkurl{10.18653/v1/P19-1212}}


\bibitem[Su et~al\mbox{.}(2023)]%
        {su2023one}
\bibfield{author}{\bibinfo{person}{Hongjin Su}, \bibinfo{person}{Weijia Shi}, \bibinfo{person}{Jungo Kasai}, \bibinfo{person}{Yizhong Wang}, \bibinfo{person}{Yushi Hu}, \bibinfo{person}{Mari Ostendorf}, \bibinfo{person}{Wen-tau Yih}, \bibinfo{person}{Noah~A Smith}, \bibinfo{person}{Luke Zettlemoyer}, {and} \bibinfo{person}{Tao Yu}.} \bibinfo{year}{2023}\natexlab{}.
\newblock \showarticletitle{One Embedder, Any Task: Instruction-Finetuned Text Embeddings}. In \bibinfo{booktitle}{\emph{Annual Meeting of the Association for Computational Linguistics-ACL 2023 (09/07/2023-14/07/2023,,, Toronto, Canada)}}.
\newblock


\bibitem[Team et~al\mbox{.}(2024)]%
        {grattafiori2024llama3herdmodels}
\bibfield{author}{\bibinfo{person}{The Llama~3 Team}, \bibinfo{person}{Aaron Grattafiori}, \bibinfo{person}{Abhimanyu Dubey}, \bibinfo{person}{Abhinav Jauhri}, \bibinfo{person}{Abhinav Pandey}, \bibinfo{person}{Abhishek Kadian}, \bibinfo{person}{Ahmad Al-Dahle}, \bibinfo{person}{Aiesha Letman}, \bibinfo{person}{Akhil Mathur}, \bibinfo{person}{Alan Schelten}, \bibinfo{person}{Alex Vaughan}, \bibinfo{person}{Amy Yang}, \bibinfo{person}{Angela Fan}, \bibinfo{person}{Anirudh Goyal}, \bibinfo{person}{Anthony Hartshorn}, \bibinfo{person}{Aobo Yang}, \bibinfo{person}{Archi Mitra}, \bibinfo{person}{Archie Sravankumar}, \bibinfo{person}{Artem Korenev}, \bibinfo{person}{Arthur Hinsvark}, \bibinfo{person}{Arun Rao}, \bibinfo{person}{Aston Zhang}, \bibinfo{person}{Aurelien Rodriguez}, \bibinfo{person}{Austen Gregerson}, \bibinfo{person}{Ava Spataru}, \bibinfo{person}{Baptiste Roziere}, \bibinfo{person}{Bethany Biron}, \bibinfo{person}{Binh Tang}, \bibinfo{person}{Bobbie Chern}, \bibinfo{person}{Charlotte Caucheteux},
  \bibinfo{person}{Chaya Nayak}, \bibinfo{person}{Chloe Bi}, \bibinfo{person}{Chris Marra}, {and} \bibinfo{person}{Chris McConnell}.} \bibinfo{year}{2024}\natexlab{}.
\newblock \bibinfo{title}{The Llama 3 Herd of Models}.
\newblock
\showeprint[arxiv]{2407.21783}~[cs.AI]
\urldef\tempurl%
\url{https://arxiv.org/abs/2407.21783}
\showURL{%
\tempurl}


\bibitem[VM et~al\mbox{.}(2024)]%
        {vm2024finetune-cost}
\bibfield{author}{\bibinfo{person}{Kushala VM}, \bibinfo{person}{Harikrishna Warrier}, \bibinfo{person}{Yogesh Gupta}, {et~al\mbox{.}}} \bibinfo{year}{2024}\natexlab{}.
\newblock \showarticletitle{Fine tuning llm for enterprise: Practical guidelines and recommendations}.
\newblock \bibinfo{journal}{\emph{arXiv preprint arXiv:2404.10779}} (\bibinfo{year}{2024}).
\newblock


\bibitem[Wang et~al\mbox{.}(2024a)]%
        {wang2024textembeddingsweaklysupervisedcontrastive}
\bibfield{author}{\bibinfo{person}{Liang Wang}, \bibinfo{person}{Nan Yang}, \bibinfo{person}{Xiaolong Huang}, \bibinfo{person}{Binxing Jiao}, \bibinfo{person}{Linjun Yang}, \bibinfo{person}{Daxin Jiang}, \bibinfo{person}{Rangan Majumder}, {and} \bibinfo{person}{Furu Wei}.} \bibinfo{year}{2024}\natexlab{a}.
\newblock \bibinfo{title}{Text Embeddings by Weakly-Supervised Contrastive Pre-training}.
\newblock
\showeprint[arxiv]{2212.03533}~[cs.CL]
\urldef\tempurl%
\url{https://arxiv.org/abs/2212.03533}
\showURL{%
\tempurl}


\bibitem[Wang et~al\mbox{.}(2024b)]%
        {wang2024multilinguale5textembeddings}
\bibfield{author}{\bibinfo{person}{Liang Wang}, \bibinfo{person}{Nan Yang}, \bibinfo{person}{Xiaolong Huang}, \bibinfo{person}{Linjun Yang}, \bibinfo{person}{Rangan Majumder}, {and} \bibinfo{person}{Furu Wei}.} \bibinfo{year}{2024}\natexlab{b}.
\newblock \bibinfo{title}{Multilingual E5 Text Embeddings: A Technical Report}.
\newblock
\showeprint[arxiv]{2402.05672}~[cs.CL]
\urldef\tempurl%
\url{https://arxiv.org/abs/2402.05672}
\showURL{%
\tempurl}


\bibitem[Yang et~al\mbox{.}(2022)]%
        {yang2022optimizing}
\bibfield{author}{\bibinfo{person}{Zhihui Yang}, \bibinfo{person}{Zuozhi Wang}, \bibinfo{person}{Yicong Huang}, \bibinfo{person}{Yao Lu}, \bibinfo{person}{Chen Li}, {and} \bibinfo{person}{X~Sean Wang}.} \bibinfo{year}{2022}\natexlab{}.
\newblock \showarticletitle{Optimizing machine learning inference queries with correlative proxy models}.
\newblock \bibinfo{journal}{\emph{Proceedings of the VLDB Endowment}} \bibinfo{volume}{15}, \bibinfo{number}{10} (\bibinfo{year}{2022}), \bibinfo{pages}{2032--2044}.
\newblock


\bibitem[Zeighami et~al\mbox{.}(2025)]%
        {10.1145/3769776}
\bibfield{author}{\bibinfo{person}{Sepanta Zeighami}, \bibinfo{person}{Shreya Shankar}, {and} \bibinfo{person}{Aditya Parameswaran}.} \bibinfo{year}{2025}\natexlab{}.
\newblock \showarticletitle{Cut Costs, Not Accuracy: LLM-Powered Data Processing with Guarantees}.
\newblock \bibinfo{journal}{\emph{Proc. ACM Manag. Data}} \bibinfo{volume}{3}, \bibinfo{number}{6}, Article \bibinfo{articleno}{311} (\bibinfo{date}{Dec.} \bibinfo{year}{2025}), \bibinfo{numpages}{26}~pages.
\newblock
\href{https://doi.org/10.1145/3769776}{doi:\nolinkurl{10.1145/3769776}}


\bibitem[Zhang et~al\mbox{.}(2025)]%
        {10.1145/3725352}
\bibfield{author}{\bibinfo{person}{Enhao Zhang}, \bibinfo{person}{Nicole Sullivan}, \bibinfo{person}{Brandon Haynes}, \bibinfo{person}{Ranjay Krishna}, {and} \bibinfo{person}{Magdalena Balazinska}.} \bibinfo{year}{2025}\natexlab{}.
\newblock \showarticletitle{Self-Enhancing Video Data Management System for Compositional Events with Large Language Models}.
\newblock \bibinfo{journal}{\emph{Proc. ACM Manag. Data}} \bibinfo{volume}{3}, \bibinfo{number}{3}, Article \bibinfo{articleno}{215} (\bibinfo{date}{June} \bibinfo{year}{2025}), \bibinfo{numpages}{29}~pages.
\newblock
\href{https://doi.org/10.1145/3725352}{doi:\nolinkurl{10.1145/3725352}}


\bibitem[Zhang et~al\mbox{.}(2024)]%
        {zhang2024data}
\bibfield{author}{\bibinfo{person}{Shuo Zhang}, \bibinfo{person}{Zezhou Huang}, {and} \bibinfo{person}{Eugene Wu}.} \bibinfo{year}{2024}\natexlab{}.
\newblock \showarticletitle{Data cleaning using large language models}.
\newblock \bibinfo{journal}{\emph{arXiv preprint arXiv:2410.15547}} (\bibinfo{year}{2024}).
\newblock


\bibitem[Zheng et~al\mbox{.}(2024)]%
        {zheng2024sglang}
\bibfield{author}{\bibinfo{person}{Lianmin Zheng}, \bibinfo{person}{Liangsheng Yin}, \bibinfo{person}{Zhiqiang Xie}, \bibinfo{person}{Chuyue~Livia Sun}, \bibinfo{person}{Jeff Huang}, \bibinfo{person}{Cody~Hao Yu}, \bibinfo{person}{Shiyi Cao}, \bibinfo{person}{Christos Kozyrakis}, \bibinfo{person}{Ion Stoica}, \bibinfo{person}{Joseph~E Gonzalez}, {et~al\mbox{.}}} \bibinfo{year}{2024}\natexlab{}.
\newblock \showarticletitle{Sglang: Efficient execution of structured language model programs}.
\newblock \bibinfo{journal}{\emph{Advances in neural information processing systems}}  \bibinfo{volume}{37} (\bibinfo{year}{2024}), \bibinfo{pages}{62557--62583}.
\newblock


\bibitem[Zhong et~al\mbox{.}(2024)]%
        {10.5555/3691938.3691949}
\bibfield{author}{\bibinfo{person}{Yinmin Zhong}, \bibinfo{person}{Shengyu Liu}, \bibinfo{person}{Junda Chen}, \bibinfo{person}{Jianbo Hu}, \bibinfo{person}{Yibo Zhu}, \bibinfo{person}{Xuanzhe Liu}, \bibinfo{person}{Xin Jin}, {and} \bibinfo{person}{Hao Zhang}.} \bibinfo{year}{2024}\natexlab{}.
\newblock \showarticletitle{DistServe: disaggregating prefill and decoding for goodput-optimized large language model serving}. In \bibinfo{booktitle}{\emph{Proceedings of the 18th USENIX Conference on Operating Systems Design and Implementation}} (Santa Clara, CA, USA) \emph{(\bibinfo{series}{OSDI'24})}. \bibinfo{publisher}{USENIX Association}, \bibinfo{address}{USA}, Article \bibinfo{articleno}{11}, \bibinfo{numpages}{18}~pages.
\newblock
\showISBNx{978-1-939133-40-3}


\end{thebibliography}
\end{multicols}

\end{document}